\documentclass[twocolumn,epjc3]{svjour3}  

\usepackage[T1]{fontenc} 
\usepackage{epsfig}  
\usepackage{graphicx}  
\usepackage{amssymb}
\usepackage{amsmath}
\usepackage{pifont}
\usepackage[english]{babel}
\usepackage{color}  
\usepackage{multirow}
\definecolor{darkgreen}{rgb}{0.2,0.5, 0.2}

\usepackage{dcolumn,booktabs}
\newcolumntype{d}[1]{D{.}{.}{#1}}
\newcommand\mc[1]{\multicolumn{1}{c}{#1}} 

\usepackage{lipsum}

\definecolor{mbscolor}{rgb}{0.60, 0.0, 0.65}

\usepackage{dsfont}

\begin{document}

\title{Skyrme-Hartree-Fock-Bogoliubov mass models on a 3D mesh: 
       II. time-reversal symmetry breaking.}

\author{ Wouter Ryssens  \thanksref{e1,addr1}
        \and
        Guillaume Scamps \thanksref{addr1,addr2}
        \and
        Stephane Goriely \thanksref{addr1}
        \and
        Michael Bender   \thanksref{addr3}    
}
\thankstext{e1}{e-mail: wouter.ryssens@ulb.be}

\institute{Institut d'Astronomie et d'Astrophysique, Universit\'e Libre de Bruxelles, Campus de la Plaine CP 226, 1050 Brussels, Belgium \label{addr1}
           \and
           Department of Physics, University of Washington, Seattle, Washington 98195-1560, USA \label{addr2}
           \and
           Universit{\'e} de Lyon, Universit{\'e} Claude Bernard Lyon 1, CNRS, IP2I Lyon / IN2P3, UMR 5822, F-69622, Villeurbanne, France \label{addr3}
}

\maketitle

\begin{abstract} 

Models based on nuclear energy density functionals can provide access to a 
multitude of observables for thousands of nuclei in a single framework with 
microscopic foundations. Such models can rival the accuracy of more 
phenomenological approaches, but doing so requires adjusting parameters to thousands
of nuclear masses. To keep such large-scale fits feasible, several symmetry restrictions 
are generally imposed on the nuclear configurations. One such example is
time-reversal invariance, which is generally enforced via 
the Equal Filling Approximation (EFA). Here we lift this assumption, enabling us 
to access the spin and current densities in the ground states of odd-mass and 
odd-odd nuclei and which contribute to the total energy of such nuclei through 
so-called `time-odd' terms. We present here the Skyrme-based BSkG2 model whose 
parameters were adjusted to essentially all known nuclear masses without relying
on the EFA, refining our earlier work [G. Scamps et al., EPJA 57, 333 (2021)]. 
Moving beyond ground state properties, we also incorporated information 
on the fission barriers of actinide nuclei in the parameter adjustment. 
The resulting model achieves a root-mean-square (rms) deviation of (i) 0.678 MeV
on 2457 known masses, (ii) 0.027 fm on 884 measured charge radii, (iii) 
0.44 MeV and 0.47 MeV, respectively, on 45 reference values for primary and 
secondary fission barriers of actinide nuclei, and (iv) 0.49 MeV on 28 fission 
isomer excitation energies. We limit ourselves here to a description of the 
model and the study the impact of lifting the EFA on ground state properties such as 
binding energies, deformation and pairing, deferring a detailed discussion of 
fission to a forthcoming paper. 

\end{abstract}

\section{Introduction}

Nucleosynthesis simulations, and in particular those dealing with the r-process,
require nuclear data across the nuclear chart for several different 
quantities~\cite{Arnould20}. For the overwhelming majority of nuclei, 
experimental information on observables such as the binding energy is 
unavailable due to the difficulty of synthesising and detecting short-lived 
neutron-rich nuclei in accelerators and laboratories. Even for nuclei close to 
stability, more complex quantities such as nuclear level densities and the 
$\gamma$-ray strength function can be either difficult or just very expensive 
to measure.

Nuclear theory needs to fill this knowledge gap by constructing models
capable of providing reliable extrapolations for nuclei at the extremes of 
energy, angular momentum and isospin. The complexity of the nuclear many-body 
problem implies that such systematic modelling of multiple thousands of 
(mostly heavy) nuclei cannot be tackled from first principles, despite the 
recent successes of ab initio approaches~\cite{Hergert20}. 
Some degree of phenomenology in a global nuclear model is thus unavoidable, 
but too much of it negatively impacts the reliability of extrapolations to 
exotic nuclei.

Methods based on energy density functionals (EDFs) provide an attractive compromise:
they provide access to many different quantities of interest in a 
consistent framework with microscopic foundations, while calculations 
for thousands of nuclei remain feasible~\cite{Bender03,Goriely16}. 
A nuclear EDF links the energy of a nucleus to the precise configuration
of its nucleonic densities through coupling constants, which are the main source
of phenomenology in these approaches. Theory offers little \emph{a priori} 
information on the values of these (often numerous) constants, such that they 
need to be fitted to 
experimental data. Many different fitting strategies for several 
different types of EDFs have been suggested in the literature~\cite{Dutra12,Dutra14}, 
aimed at producing models with differing goals and regions of applicability.

If furnishing data to applications is the goal, it is natural to 
impose an excellent \emph{global} reproduction of nuclear masses, as 
these set the energy scales involved in nuclear reactions and decays. 
The Brussels-Montréal (BSk) models, based on EDFs of the Skyrme
type\footnote{Global mass models based on other types of EDF exist, 
but their parameter adjustment has not been pushed as far. See Ref.~\cite{Goriely09b} 
for a model based on an EDF of the Gogny type and Ref.~\cite{Pena16} for 
a relativistic mean-field model.}, have shown that incorporating
essentially all known masses in the fit protocol
leads to an excellent description of masses. The members of the BSk-family
achieve a root-mean-square (rms) deviation typically below 800 keV, and often 
close to 500 keV~\cite{Goriely16,Goriely13a}, which is competitive with 
the global nuclear models that have less microscopic foundations~\cite{Moller16}. 
Furthermore, this accuracy for the masses is combined with 
(i) a global description of other properties of finite nuclei 
such as charge radii and fission barriers, and (ii) a description of
infinite nuclear matter that is generally comparable to ab-initio calculations.

While central to the success of the strategy, the inclusion of thousands of  
masses renders the fit computationally demanding. Out of necessity, nuclei are 
thus typically not treated in the most general way possible during the parameter
adjustment. For the BSk-family, the computational demands resulted in two
practical restrictions: the nuclear configurations were (i) highly 
restricted by symmetry assumptions and (ii) numerically represented through an 
expansion of the single-particle wavefunctions in a limited number of harmonic 
oscillator states. These two restrictions greatly lower the computational 
complexity of a fit, but the first restriction limits the generality of 
the resulting model while the second limits its numerical accuracy.

To provide nuclear data free of these limitations, we developed in 
Ref.~\cite{Scamps21} the first Brussels-Skyrme-on-a-Grid model, BSkG1. 
This model was adjusted using the MOCCa code~\cite{Ryssens16}, which relies on a
numerical representation of the nucleus in three-dimensional coordinate space.
Such a representation allows for an improved control of numerical accuracy~\cite{Ryssens15b}, 
while imposing less stringent symmetry conditions. Where the BSk models 
considered all nuclei to be axially symmetric, the BSkG1 model offers a more 
general description of nuclear ground states by allowing for triaxial deformation.
It retains the overall quality of the older models in terms of the global
description of masses, charge radii and infinite nuclear matter properties.
BSkG1 was but a first step in this direction, as its parameter 
adjustment was still limited in a number of ways~\cite{Scamps21}. First, 
the nuclear configurations allowed for were not the most general possible, but
restricted by the assumption of time-reversal invariance and reflection symmetry.
Secondly, we restricted our study to nuclear ground states, ignoring the 
quality of the model with respect to other properties of atomic nuclei that 
impact astrophysical applications. 

Allowing for reflection-asymmetric shapes is not a priority for a 
description of masses, as in a mean-field approach octupole correlations are 
known to impact nuclear ground states for only a very limited
number of nuclei in a few isolated regions of the nuclear chart 
\cite{Robledo11,Chen21}.
The situation is very different for odd-mass and odd-odd nuclei which constitute
roughly three quarters of the chart of nuclei and whose ground states all have 
finite angular momentum, and hence break time-reversal symmetry. An internally
consistent EDF-based description of such nuclei requires a many-body state that 
has non-vanishing time-odd spin and current densities. Through the so-called 
\emph{time-odd} terms of the EDF~\cite{Duguet2001a,Hellemans12}, these densities
also contribute to the binding energy of the nucleus.

In fact, self-consistent mean-field calculations of the ground states of
even-even nuclei are among the very few exceptional cases for which breaking of 
time-reversal symmetry and the consideration of time-odd terms in the EDF is 
\textit{not} necessary.
For instance, time-odd terms appear naturally in methods to describe nuclear 
dynamics, such as time-dependent mean-field~\cite{Maruhn06} and RPA
approaches~\cite{Bender02,Kortelainen15}. They also contribute
in beyond-mean-field approaches like the restoration of spatial symmetries and the 
mixing of states with different shapes in the Generator Coordinate Method (GCM)
\cite{Bonche90} or Adiabatic time-dependent HFB 
\cite{Hinohara06a,Kluepfel08a,Petrik18,Washiyama21a}, although in most other 
published work on the latter the time-odd terms are usually neglected.
The vast majority of excited
states of all nuclei, including even-even ones, exhibit non-zero
angular momentum and are affected as well, independently of the nature of the
excitation: the description of rotational states through self-consistent 
cranking calculations introduces an external field that breaks
time-reversal symmetry~\cite{Bonche87,Dobaczewski95}, 
while the modelling of non-collective states breaks time-reversal through the 
creation of one or more quasiparticle excitations~\cite{Robledo14}.
Degrees of freedom that break time-reversal invariance affect other observables
beyond masses, such as magnetic moments~\cite{Sassarini21}. 

The relevance of the time-odd terms is also not limited to finite nuclei: the
time-odd spin-spin interaction terms contribute to the equation of state of spin- 
and spin-isospin polarised infinite homogeneous nuclear matter. Their contribution
can be expected to influence the composition of neutron star crusts in 
the presence of strong magnetic fields~\cite{Pena11}, and even induce a phase 
transition to homogeneous polarised matter~\cite{Margueron02,Chamel09,Chamel10b,Cao10}. 
Similarly, gradients of spin densities in the EDF can induce finite-size 
instabilities that lead to a phase transition to inhomogeneous infinite polarised 
matter \cite{Hellemans12,Pastore13,Pastore15}.
The absence of indications of such phase transitions in finite nuclei 
and stellar objects however limits the range of possible coupling constants for the 
spin terms in the EDF.

Despite their relevance to the properties of finite nuclei and infinite matter, 
systematic studies of the time-odd channel of EDFs are scarce. Truly global 
approaches that study its impact in finite nuclei up to the drip lines 
are to the best of our knowledge non-existent\footnote{
Ref.~\cite{Pototzky10} comes close to a truly global study for 
three different Skyrme parameterizations, but considers no odd-odd nuclei. 
Ref.~\cite{Afanasjev10a} presents detailed results limited to a large but 
incomplete selection of isotopic and isotonic chains for different 
relativistic models.}. Most large-scale calculations employ the equal 
filling approximation (EFA) instead, which, eliminates all 
contributions of the time-odd terms by construction~\cite{Perez08}. This approximation is
popular for two reasons: (i) the absence of consensus in the literature on the 
optimal treatment of the time-odd terms, at least for EDFs of the Skyrme type 
and (ii) the complexity of solving the EDF equations for odd-mass and odd-odd 
nuclei. The technical complexity is related to practical considerations about
 CPU time and memory, but also due to the difficulty of 
reliably converging such calculations~\cite{Schunck10}.

We present here the next entry in the BSkG-series, BSkG2, which moves beyond the 
limitations of its predecessor in two ways. First, we allow for more 
general nuclear configurations by lifting the assumption of time-reversal 
invariance, allowing for finite angular momentum in the nuclear ground state
and exploring for the first time the influence of time-odd terms on the 
properties of finite nuclei in the context of a global EDF-based model.
The resulting model offers a description of odd-mass and odd-odd nuclei that is 
on par with that of even-even nuclei, i.e. without invoking the EFA.

Secondly, we incorporate information on the fission barriers of actinides 
in the parameter readjustment in order to control the properties of the model
at large deformation. Indeed, fitting a model to nuclear masses does 
not explore all regions of deformation space, such that a good mass fit 
will not necessarily guarantee accurate fission barriers. As in
Ref.~\cite{Goriely07}, we show it is possible to obtain drastically improved 
fission barriers while retaining the quality of the mass fit, through 
the adjustment of a phenomenological vibrational term that we did not include in
the earlier BSkG1 fit. Our calculation of fission barriers, performed to 
adjust this vibrational term, included octupole and triaxial degrees of freedom
for systems with both even and off numbers of nucleons. Due to this complexity, 
we postpone discussion of all our material related to fission to a forthcoming 
paper~\cite{Ryssens22}. Here, we limit our discussion to the construction 
of the BSkG2 model and its description of both the ground state of finite nuclei
and infinite nuclear matter. 

This paper is organized as follows: we start by explaining the ingredients
of the mass-model as well as the technical aspects of the calculations and
parameter adjustment in Sec.~\ref{sec:massmodel}. In Sec.~\ref{sec:model}, 
we discuss the models ground state and infinite nuclear matter properties. 
Our conclusions and outlook are presented in Sec.~\ref{sec:conclusions}.


\section{Construction of the mass model}
\label{sec:massmodel}

\subsection{The nuclear mass}
\label{Sec:Energy} 
To describe an atomic nucleus, we employ an auxiliary state of the 
Bogoliubov type: $|\Phi \rangle$. As for the earlier BSkG1 model~\cite{Scamps21}, 
we define the total energy $E_{\rm tot}$ of this state:
\begin{align}
E_{\rm tot} &= E_{\rm HFB} + E_{\rm corr} \, . 
\label{eq:Etot}
\end{align}
The total energy consists of two parts: $E_{\rm HFB}$ is the self-consistent mean-field 
energy, while $E_{\rm corr}$ is a set of perturbative corrections. 
The total energy $E_{\rm tot}$ models \emph{minus} the nuclear binding energy,
$B_{\text{nuc}}(Z,N) = - E_{\rm tot}$, of a nucleus
with mass number $A$, composed out of $Z$ protons and $N$ neutrons. 
Mass tables, however, list \emph{atomic masses}, which include $Z$ bound 
electrons. In units of MeV, these are given by
\begin{equation}
M(N,Z) = E_{\rm tot} + N M_n + Z (M_p + M_e) - B_{\rm el.}(Z) \, ,
\label{eq:mass_def}
\end{equation}
where $M_n$ and $M_p$ are the masses of bare nucleons, $M_e$ is the mass of 
an electron and $B_{\rm el.}(Z)$ is a simple analytical estimate for the binding
energy of the electrons~\cite{Lunney03,Wan21}. The mass excess can then 
be obtained by subtracting $A$ times the atomic mass
unit from Eq.~\eqref{eq:mass_def}.

The mean-field energy is constructed from five contributions
\begin{align}
E_{\rm HFB} &= E_{\rm kin} + E_{\rm Sk} + E_{\rm pair} + E_{\rm Coul}  + E^{(1)}_{\rm cm}\, ,
\label{eq:Ehfb}
\end{align}
which are, respectively, the contributions of the kinetic energy,
the Skyrme effective interaction,
a zero-range pairing interaction with appropriate cutoffs~\cite{Krieger90}, 
the Coulomb force~\cite{Goriely08}, and the one-body part of 
the centre-of-mass correction~\cite{Bender00}. 
The correction energy $E_{\rm corr}$ consists of four parts:
\begin{align}
E_{\rm corr}&= E_{\rm rot} + E_{\rm vib} + E^{(2)}_{\rm cm} + E_{\rm W}  \, ,
\label{eq:Ecorr}
\end{align}
which are, respectively, the rotational correction~\cite{Tondeur00}, 
the vibrational correction~\cite{Goriely13b}, the two-body part of the 
centre-of-mass correction~\cite{Bender00}, and the Wigner energy~\cite{Goriely03}. 

Most ingredients of the total energy are identical to those
we employed in the BSkG1 model; the only exceptions are (i) the 
time-odd contributions to the Skyrme energy $E_{\rm Sk}$ and pairing energy 
$E_{\rm pair}$ and (ii) the inclusion of the vibrational correction. 
We will discuss these in detail below; for our treatment of all other terms, 
we refer the reader to Ref.~\cite{Scamps21}.

\subsubsection{Time-odd densities and terms}
\label{sec:SkyrmeEnergy}
We write the Skyrme energy as an integral over an energy density, composed out 
of four parts $\mathcal{E}_{t, \rm e/o}$:
\begin{align}
\label{eq:Eskyrme}
E_{\rm Sk} &= \int d^3 \bold{r} \sum_{t = 0,1} \left[ 
                                             \mathcal{E}_{t, \rm e}(\bold{r})
                                          + \mathcal{E}_{t, \rm o} (\bold{r}) 
                                          \right]
                                          \, ,
\end{align}
where $t=0,1$ is an isospin index. All contributions to the Skyrme energy are
constructed from a set of five local one-body densities that characterize 
the auxiliary state $|\Phi\rangle$. Three of these already figured in Ref.~\cite{Scamps21}: 
$\rho_t(\bold{r}), \tau_t(\bold{r})$ and $\bold{J}_t(\bold{r})$. We employ two 
additional densities, $\bold{s}_t(\bold{r})$ and $\bold{j}_t(\bold{r})$, 
which are the spin and current density respectively. Their definition in terms 
of the single-particle wavefunctions are standard in the literature and can be 
found for instance in Ref.~\cite{Ryssens19}.

All five densities transform in a straightforward way under time-reversal. 
Using a superscript $T$ to indicate a density characterizing the time-reversed
auxiliary state $\check{T}|\Phi \rangle$, we have
\begin{subequations}
\begin{align}
\rho_t^T (\bold{r})     &= + \rho_t(\bold{r}) \, , \\
\tau_t^T (\bold{r})     &= + \tau_t(\bold{r}) \, , \\
\bold{J}_t^T (\bold{r}) &= + \bold{J}_t(\bold{r}) \, , \\
\bold{s}_t^T (\bold{r}) &= - \bold{s}_t(\bold{r}) \, , \label{eq:sTodd} \\
\bold{j}_t^T (\bold{r}) &= - \bold{j}_t(\bold{r}) \, . \label{eq:jTodd}
\end{align}
\end{subequations}
The three densities $\rho_t(\bold{r}), \tau_t(\bold{r})$ and $\bold{J}(\bold{r})$
are time-even, while $\bold{s}_t(\bold{r})$ and $\bold{j}(\bold{r})$ are time-odd.
If we assume time-reversal invariance, meaning that the time-reversed auxiliary state
is equal to the auxiliary state up to a phase, then Eqs.~\eqref{eq:sTodd} and 
\eqref{eq:jTodd} imply that the spin and current density vanish identically.

The Skyrme energy densities $\mathcal{E}_{t, \rm e}(\bold{r})$ and 
$\mathcal{E}_{t, \rm o}(\bold{r})$ of Eq.~\eqref{eq:Eskyrme} are built from
suitable bilinear combinations of either time-even or time-odd local densities.
The energy densities themselves are necessarily even under time-reversal, 
but we will follow common practice in the literature and refer to them 
as \textit{time-even} and \textit{time-odd} energy densities, respectively. 

The time-even energy densities are given by
\begin{align}
\label{eq:Skyrme_te}
\mathcal{E}_{t, \rm e}(\bold{r})
& =  
              C^{\rho\rho}_t \, \rho_t^2 (\bold{r})
            + C^{\rho\rho\rho^{\gamma}}_t \rho_0^\gamma (\bold{r}) \, \rho_t^2 (\bold{r}) \nonumber \\
&           + C^{\rho\tau}_t \, \rho_t (\bold{r}) \, \tau_t (\bold{r})
            + C^{\rho \Delta \rho}_t \, \rho_t (\bold{r}) \, \Delta \rho_t (\bold{r})  \nonumber \\
&           + C^{\rho \nabla \cdot J}_t  \rho_t (\bold{r}) \, \boldsymbol{\nabla} \cdot \bold{J}_t (\bold{r}) \, , 
\end{align}
which are identical to the ones of Ref.~\cite{Scamps21}. The 
set of ten time-even coupling constants $\{C_t\}$ consists of
combinations of ten model parameters: $t_{0-3}, x_{0-3}, W_0$ and $W_0'$
\cite{Scamps21}.

The expression of Eq.~\eqref{eq:Skyrme_te} is very similar, but not 
identical, to the energy density that can be obtained as expectation value of a 
density-dependent two-body Skyrme interaction. The differences between the EDF 
used by us and such an interaction-generated functional are (i) the absence of 
any term bilinear in the spin-current 
density $J_{t,\mu\nu}(\bold{r})$, and (ii) the inclusion of a generalized 
spin-orbit term involving a second spin-orbit parameter $W_{0}'$ as in the 
Skyrme models of Refs.~\cite{Sharma95,Reinhard95}. For the reasons recalled in
Ref.~\cite{Lesinski07}, the former is common practice for many widely-used
parameterizations of the Skyrme EDF, while the latter has become popular in recent
parameter fits as it allows for better fine-tuning of spin-orbit effects by relaxing 
the relation $W_{0} = W_{0}'$ of the standard spin-orbit interaction and thereby
leads to linearly independent coefficients $C^{\rho \nabla \cdot J}_0$ and 
$C^{\rho \nabla \cdot J}_1 \neq 3 \,  C^{\rho \nabla \cdot J}_0$ of the EDF.  The 
time-even energy density of Eq.~\eqref{eq:Skyrme_te} is similar to the one of 
Refs.~\cite{Klupfel09,Kortelainen10,Kortelainen14} and 
consists of a subset of the terms
employed by later BSk models~\cite{Goriely16,Goriely15}, omitting, however,
the density-dependence of the $\rho_t(\bold{r}) \tau_t(\bold{r})$ and 
$\rho_t(\bold{r}) \Delta \rho_t(\bold{r})$ terms introduced in Ref.~\cite{Chamel09}.
A parameter fit of the more complete EDF of~\cite{Goriely16,Goriely15} 
with the tools described here will be the subject of future work.

There is less consensus in the literature on the time-odd part of the 
functional, and different strategies to set-up the functional form 
in the time-odd channel and to determine its coupling constants have
been proposed in the literature~\cite{Hellemans12,Bender02,Pototzky10,Schunck10}. 
Motivated both by 
physical and practical arguments, we employ the following expression
\begin{align}
\label{eq:Skyrme_to}
\mathcal{E}_{t, \rm o}(\bold{r})
& =  
              C^{s s}_t \, \bold{s}_t (\bold{r}) \cdot \bold{s}_t (\bold{r})
            + C^{s s \rho^{\gamma}}_t \rho_0^\gamma (\bold{r}) \, \bold{s}_t (\bold{r}) \cdot \bold{s}_t (\bold{r}) \nonumber \\
&           + C^{j j}_t \, \bold{j}_t (\bold{r}) \cdot \bold{j}_t (\bold{r}) 
            + C^{j \nabla s}_t \, \bold{j}_t (\bold{r}) \cdot \nabla \times \bold{s}_t (\bold{r}) 
\, .
\end{align}
We will refer to the terms on the first line collectively as \emph{spin} terms, 
while the second line is composed of the \emph{current terms} and the \emph{time-odd 
spin-orbit terms}. The expression of Eq.~\eqref{eq:Skyrme_to} is again close to 
the form generated by a density-dependent two-body Skyrme interaction, with 
modifications that (i) guarantee Galilean invariance for the EDF as a whole and 
(ii) eliminate the major source of spurious finite-size instabilities. 

Galilean invariance of a non-relativistic EDF ensures that the binding energy 
is the same in all inertial frames. Without respecting this invariance, one cannot do
meaningful calculations of nuclear dynamics. In order to impose this 
symmetry~\cite{Dobaczewski95}, we (i) do not include terms of the form 
$\bold{s}_t(\bold{r}) \cdot \bold{T}_t(\bold{r})$ in Eq.~\eqref{eq:Skyrme_to}, 
which reflects our choice to omit terms that are bilinear in the spin-current 
tensor density $J_{t,\mu\nu}(\bold{r})$ in the time-even energy density 
\cite{Hellemans12,Lesinski07}, and; (ii) we set $C^{jj}_t = - C^{\rho \tau}_t$ 
and $C^{j \nabla s}_t = C^{\rho \nabla \cdot J}_t$~\cite{Dobaczewski95}.
Taking Eq.~\eqref{eq:Skyrme_te} as the time-even energy density and imposing
Galilean invariance, determines all coupling constants in Eq.~\eqref{eq:Skyrme_to}
with the exception of those of the spin terms, $C^{ss}_t$ and $C^{ss \rho^\gamma}_t$.

We have not included terms of the form\footnote{We note that including or dropping 
these terms does not affect the overall Galilean invariance of the EDF, 
as they are Galilean invariant by themselves~\cite{Dobaczewski95}.} 
$\bold{s}_t(\bold{r}) \cdot \Delta \bold{s}_t(\bold{r})$.
For large positive values of the coupling constants $C^{s \Delta s}_t$
of these terms, their contribution leads to diverging calculations as they
give rise to  spurious finite-size spin instabilities of nuclear
matter~\cite{Hellemans12,Pastore13,Davesne15} in spite of their contribution to
the total binding energy remaining on the order of a few hundreds of keV when the
instability sets in~\cite{Hellemans13}. Such instabilities in either the 
spin or spin-isospin channel are found for the vast majority of Skyrme 
interactions\footnote{Such instabilities can arise 
as well for EDFs of the Gogny type~\cite{Davesne21}, although for existing 
parameterizations of the Gogny interaction it is the isospin channel that 
is the most problematic~\cite{Martini19,Gonza21,Davesne21} instead of the spin 
and spin-isospin channels that are unstable for many Skyrme parameterizations
when $C^{s \Delta s}_t$ is calculated through the usual relations
from $t_1$, $x_1$, $t_2$, and $x_2$~\cite{Hellemans12}.
}, but they can be artificially hidden by either imposing symmetry 
restrictions on the auxiliary many-body state or by using an inappropriately 
coarse numerical representation~\cite{Hellemans13}. In a three-dimensional coordinate space 
representation such as ours, such instabilities are virtually guaranteed to 
spoil the calculations if present, which would make the parameter adjustment of 
the model impossible. Techniques to avoid such instabilities during the 
adjustment process exist~\cite{Pastore13,Pastore15}, but we have opted to remove
the offending terms; this simple recipe has historically often been used to stabilise 
time-reversal breaking calculations~\cite{Hellemans12,Bonche87,Pototzky10}. 
In any case, their contribution to to ground-state energies will be 
significantly smaller than the typical size of the other time-odd terms~\cite{Hellemans12}.

While spurious instabilities due to other terms of the EDF could not be ruled 
out a priori, we have not encountered any sign of them during the parameter 
adjustment process, the production of the mass table or any other calculation 
presented here.

\subsubsection{Time-odd terms in the pairing interaction}
In the particle-particle channel, we employ the following simple pairing EDF
\begin{subequations}
\begin{align}
\label{eq:pairingenergy}
E_{\rm pair} &= \int \, d^3 \bold{r} \sum_{q=p,n} g_q(\bold{r})
\, 
\tilde{\rho}_q^*(\bold{r}) \tilde{\rho}_q(\bold{r})
\, ,  \\
g_q(\bold{r}) &= \frac{V_{\rm \pi q}}{4} 
 \left[1 - \eta \left(\frac{\rho_0(\bold{r})}{\rho_{\rm ref}}\right)^{\alpha} \right] \, ,
 \label{eq:pairingformfactor}
\end{align}
\end{subequations}
where $\tilde{\rho}_q(\bold{r})$ is the local pair density of nucleon species $q$
and we take $\rho_{\rm ref} = 0.16$ fm$^{-3}$. The parameters $\eta$ and 
$\alpha$ of the spatial form factor and the pairing strengths $V_{\pi q}$($q=p,n$) 
of each nucleon species are the adjustable parameters of this EDF that is 
formally identical to the pairing energy density employed for BSkG1 in 
Ref.~\cite{Scamps21}. We also employ the same cutoff procedure as described 
there to avoid the divergence of the zero-range pairing 
energy~\eqref{eq:pairingenergy} with increasing model space.

The energy density of Eq.~\eqref{eq:pairingenergy} can be separated into 
a time-even and time-odd energy density along the same lines as the particle-hole
channel. While not often mentioned in the literature, the pair density 
$\tilde{\rho}_q(\bold{r})$ is in general a complex function and its real and
imaginary parts are even and odd under time-reversal, 
respectively~\cite{Perlinska04}. 
The pairing energy of nuclei with odd proton and/or neutron number thus contains
a contribution due to $\text{Im} \left[ \tilde{\rho}_q(\bold{r}) \right]$
taking non-zero values. This contribution is generally small but reaches a 
few tens of keV for a handful of nuclei in our calculations.

\subsubsection{The vibrational correction}
\label{sec:vib_corr}
Another new ingredient of the model is the phenomenological vibrational correction:
\begin{subequations}
\begin{align}
\label{eq:vibrationalcorrection}
E_{\rm vib} &= 
-\sum_{\mu = x,y,z} f_{\mu, \rm vib} \frac{\langle \hat{J}_{\mu}^2 \rangle}{2 \mathcal{I}^{\rm B}_{\mu}} \, , \\
f_{\mu, \rm vib} &= d B_{\mu} e^{- l \left( B_{\mu} - B_0 \right)^2} \, ,\\
B_{\mu} &= \frac{\mathcal{I}^{\rm B}_{\mu}}{\mathcal{I}_c} \, .
\label{eq:B_compare}
\end{align}
\end{subequations}
where $d,l$ and $B_0$ are adjustable parameters and
$\mathcal{I}_c = \frac{2}{15}m R^2 A$ is one third of the MOI of a rigid rotor 
of radius $R  = 1.2 A^{1/3}$ comprised of $A$ nucleons. $\mathcal{I}^{\rm B}_{\mu}$ 
is the Belyaev moment of inertia (MOI) around the Cartesian axis $\mu = x,y,z$~\cite{Belyaev61}, 
which we discuss below in more detail in Sec.~\ref{sec:mois}.

Equation~\eqref{eq:vibrationalcorrection} has the same form as the rotational 
correction: an expectation value of the angular momentum squared 
divided by the MOI modulated by a dimensionless factor. This form was inspired 
by the vibrational correction of Ref.~\cite{Goriely07}, but generalized for 
systems without axial symmetry. It should be stressed that 
Eq.~\eqref{eq:vibrationalcorrection} only takes account of the deformation 
dependence of the vibrational correction. Since it vanishes for spherical 
nuclei, the vibrational correction for such nuclei is absorbed into the 
fitted force parameters. For a more detailed discussion of this term, 
which is crucial to our description of fission, we refer the
reader to the accompanying paper~\cite{Ryssens22}.

\subsection{Numerical treatment of ground states}
\label{sec:numerical}

\subsubsection{General comments}

In this section, we describe our treatment of nuclear ground states: how we 
represent Bogoliubov states and how we find among them the 
configuration that minimizes the total energy $E_{\rm tot}$. The core of our 
approach is identical to that described in Ref.~\cite{Scamps21}: we use the 
MOCCa code~\cite{Ryssens16} to solve the Skyrme-Hartree-Fock-Bogoliubov (HFB) 
equations on a three-dimensional Cartesian Lagrange mesh~\cite{Baye86}. We 
employ identical values for the mesh parameters ($N_x, N_y, N_z, dx$) and the 
total number of proton ($N_Z$) and neutron ($N_N$) single-particle wavefunctions
that are iterated. 

Fission barrier heights can only be obtained from potential energy surfaces,
which requires many calculations as a function of one or more collective 
coordinates. Even individual calculations on such surfaces are more costly than 
a typical ground state calculation, as coordinate meshes should be enlarged to
 accomodate very elongated shapes~\cite{Ryssens15b} and reflection symmetry cannot 
be assumed along the entire fission path. The MOCCa code offers all the tools
necessary for such calculations, whose technical aspects will be presented in 
the accompanying paper~\cite{Ryssens22}. We will thus limit the present 
discussion to those technical aspects of the treatment of ground states that 
are different from the adjustment of BSkG1 as described in Ref.~\cite{Scamps21}.

\subsubsection{Self-consistent symmetries}
\label{sec:sym}

Specific to the MOCCa code is its flexibility with respect to the 
symmetries imposed on the auxiliary state $| \Phi \rangle$. If the 
corresponding operator is linear, imposing a symmetry implies a conserved 
quantum number for the many-body state, which can be exploited to significantly 
reduce the computational cost of calculations. 

Similar to our approach in Ref.~\cite{Scamps21}, we consider configurations 
that respect three plane-reflection symmetries. The shapes of the nuclear 
density that respect these symmetries are all reflection symmetric and invariant
under rotations of $180^{\circ}$ with respect to any Cartesian axis. These 
symmetry assumptions allow us to chose single-particle states 
$|\psi_i \rangle$ ($i=1, \ldots N_N + N_Z$) with the following symmetry properties:
\begin{subequations}
\begin{align}
\label{eq:sp_zsig}
\hat{R}_z     | \psi_i \rangle &= \eta_i      | \psi_i\rangle \, , \\
\label{eq:sp_sty}
\check{S}^T_y | \psi_i \rangle &= \phantom{+} | \psi_i\rangle \, , \\
\label{eq:sp_par}
\hat{P}       | \psi_i \rangle &= p_i         | \psi_i\rangle \, ,
\end{align}
\end{subequations}
where $\hat{R}_z$ and $\check{S}^T_y \equiv \check{T} \, \hat{P} \, \hat{R}_y$ 
are the one-body $z$-signature and $y$-time-simplex operators, respectively, 
while $\hat{P}$ is the parity operator~\cite{Doba00}. 
Besides the nucleon's isospin, the eigenvalues 
$\eta_i = \pm i$ of $\hat{R}_z$ and $p_i = \pm 1$ of $\hat{P}$ are the only 
remaining conserved quantum numbers of the single-particle states. The
operator $\check{S}^T_y$ is anti-linear, a property we indicate by the inverted 
hat. As a consequence, Eq.~\eqref{eq:sp_sty} is not an eigenvalue
equation, but fixes the relative phases of the single-particle wavefunctions 
through a symmetry relation instead~\cite{Doba00}. Exploiting these spatial 
symmetries allows us to perform calculations on an effective mesh with only 
$(N_x/2, N_y/2, N_z/2)$ points, 
reducing the computational burden in memory and CPU time by a factor of eight. 
We reiterate that this set of symmetry assumptions does not allow for
reflection-asymmetric shapes: we did not explore octupole deformation 
for nuclear ground states during the parameter adjustment of BSkG2.

Time-reversal symmetry is less intuitive than the spatial symmetries in 
Eqs.~\eqref{eq:sp_zsig} - \eqref{eq:sp_par}: imposing it does not impact the 
shape of the nuclear density. Instead, conserved time-reversal symmetry
implies that spin and current densities are zero everywhere, resulting 
in vanishing expectation values of angular momentum
\begin{equation}
\langle \hat{\bold{J}} \rangle 
= \hbar \int \! d^3r \big[ \bold{r} \times \bold{j}_0(\bold{r}) 
                     + \tfrac{1}{2} \bold{s}_0(\bold{r}) \big] \, ,
\end{equation}
and other time-odd quantities. 

In addition to being antilinear, the single-particle time-reversal operator 
$\check{T}$ is also an antihermitian operator. Both properties combined 
imply that this operator does not have eigenstates and no associated 
single-particle quantum number. Instead it allows us to group states into
pairs $(k, \bar{k})$ that are connected by time-reversal with the phase convention~\cite{Wigner60}
\begin{subequations}
\begin{align}
\check{T} | \phi_k \rangle         &= + | \phi_{\bar{k}} \rangle \, , \\
\check{T} | \phi_{\bar{k}} \rangle &= - | \phi_{k} \rangle \, .
\end{align}
\end{subequations}
When $z$-signature is conserved, the members of time-reversal pairs have opposite
signature quantum number; our conventions result in 
$(\eta_k, \eta_{\bar{k}}) = (+i,-i)$.

Conserved time-reversal symmetry dictates that both members of a pair of 
single-particle states contribute with equal weights to all observables: for 
time-even quantities it implies that their contributions \textit{add}, whereas for 
time-odd quantities their contributions \textit{cancel} exactly. 
If time-reversal is conserved, it suffices to numerically represent only one 
state of each pair, such that calculations can be restricted to only $N_Z/2$ and
$N_N/2$ single-particle states. 


\subsubsection{Blocking and the EFA}
\label{sec:blocking}

Bogoliubov many-body states can be separated into four categories, following 
the number parity ($\pi_n, \pi_p$) = $(\pm 1,\pm 1)$ of each nucleon 
species~\cite{Banerjee74,RingSchuck,Bertsch09}. For a given configuration of 
the nuclear mean-field potentials, the Bogoliubov state with lowest total energy
generally has even number parity for both neutrons and protons 
\footnote{This might not hold anymore in situations with strong external 
fields~\cite{Bertsch09}. A sufficiently strong constraint on angular 
momentum as in Eq.~\eqref{eq:routhian} is one example, see Refs.~\cite{Banerjee74,BenderEDF3}.}.
This state is easy to recognize among all possible quasiparticle
vacua: it corresponds to choosing all quasiparticle energies to be strictly
positive~\cite{BenderEDF3}. We will call such a state a reference (quasiparticle) vacuum. 

For the ground states of even-even nuclei, one searches for the state with 
even proton and neutron number parity of lowest total energy: this is 
evidently a reference vacuum. For non-pathological interactions, this state is 
also time-reversal invariant: its spin and current densities vanish exactly,
resulting in vanishing expectation values of all angular momentum 
components\footnote{This is as it should be, as all observed ground 
states of even-even nuclei have spin zero.} $\hat{J}_{x/y/z}$. To not waste 
computational resources, we assumed time-reversal symmetry for 
even-even systems from the start, restricting our numerical representation to 
the $\eta_i = +i$ single-particle states for such nuclei.

For odd-mass or odd-odd nuclei, the situation is different: one requires
that the number parity of the nucleon species with odd number is
negative~\cite{BenderEDF3}. Such states can be constructed through the creation
of one or more quasiparticle excitation(s) with respect to the reference quasiparticle vacuum.
An excited quasiparticle is generally referred to as a \textit{blocked} state
in the literature~\cite{RingSchuck,Bertsch09,BenderEDF3}. For the calculations we report on here, 
these states can be labelled by the parity and $z$-signature quantum numbers $p$ and $\eta$, 
but not by an angular momentum quantum number. Independent of the quantum 
numbers of the blocked quasiparticle(s), the resulting auxiliary state will in 
general not be invariant under time-reversal symmetry; a complete calculation 
then requires representing in memory all $N_N$ and $N_Z$ states. One iteration 
in a such calculation requires twice the amount of CPU time and memory compared 
to an iteration in a time-reversal conserving one.

A simple way to sidestep the increase of computational requirements is the
EFA: this approximation replaces the pure Bogoliubov state
by a statistical mixture of two auxiliary states of odd number parity, 
which modifies the Skyrme-HFB equations only slightly~\cite{Perez08}. 
While each of these states breaks time-reversal individually, the statistical mixture of 
both does not; in this way the blocking effect of the odd neutron and/or proton
can be taken into account while retaining the computational simplicity of 
time-reversal-conserving calculations. For this reason, the EFA is very popular 
in the literature: for instance, the entire BSk-family of models~\cite{Goriely16} 
and the BSkG1 model~\cite{Scamps21} relied on it to simplify calculations. 
We mention in particular that, \emph{for time-even observables}
an EFA calculation is entirely equivalent to a complete time-reversal breaking 
calculation in the absence of time-odd terms in the EDF 
\footnote{We used this to numerically check our implementation.}. Despite this 
practical advantage, this approximation cannot be employed to study any time-odd 
quantities -- such as angular momentum or the time-odd terms of the EDF -- as they all vanish 
by construction. 

\subsubsection{Converging blocking calculations}
\label{sec:convergence}

In practice, the computational requirements are not the main obstacle to
systematic blocking calculations without the assumption of time-reversal symmetry.
Whether employing the EFA or not, the chief difficulty lies in the selection of the 
quasiparticle excitation(s) one should construct. For a given reference vacuum, 
there are usually many possible excitations with 
comparable quasiparticle energies. The need to choose between these possibilities
gives rise to two problems. The first is of a physical nature: blocking the 
quasiparticle with lowest quasiparticle energy at the current iteration
is not guaranteed to lead to the many-body state 
with lowest total energy \textit{at convergence}.
The second problem is more practical: if the blocked quasiparticle changes
dramatically from one iteration to the next, the convergence of the self-consistent procedure 
is unlikely. This problem is particularly serious (a) when many different
quasiparticles with identical quantum numbers are close in energy or (b) when 
we are dealing with light nuclei, where polarization effects can be large.

We will refer to the most widely-spread technique to solve the (Skyrme-)HFB 
equations as \emph{direct diagonalisation}: at each iteration of the self-consistent
procedure, this strategy consists of completely diagonalising the HFB 
Hamiltonian~\cite{Heenen95,Doba09,Perez17}. Using the sign of the quasiparticle
energies, one constructs the relevant reference vacuum and then uses some 
(predetermined) quasiparticle selection procedure to construct excitation(s)
on top of it. When quasiparticles can be labelled by a sufficient amount of 
quantum numbers, i.e. when multiple symmetries are imposed on the nuclear 
configuration, a consistent choice of blocked state(s) can be made at every 
iteration. In such conditions the calculation will generally converge, 
particularly when dealing with heavy, well-deformed systems. In less ideal 
conditions, the quasiparticle selection procedure can result in discontinuous 
changes in the many-body state from one iteration to the next, rendering 
convergence impossible.

The direct diagonalisation technique is the one we used to construct the BSkG1 
model using the EFA~\cite{Scamps21}. Initial efforts to employ the same 
numerical technique without the EFA failed systematically: the polarising 
effects of time-odd terms on the nucleus render convergence of 
the self-consistent procedure much more difficult\footnote{We note that 
converging calculations for systems with finite spin densities is also 
considered very difficult in the condensed matter community~\cite{Woods19}.}.
To solve this issue, we developed an approach based on the \emph{gradient method} 
to solve the HFB equations~\cite{RingSchuck,Egido95,Robledo11b}. This technique
relies on Thouless transformations to update the auxiliary state, thus ensuring 
a continuous connection between many-body states from one iteration to the next. 
We have implemented this approach in the MOCCa code, 
taking care to retain its stability in an approach relying on the two-basis 
method~\cite{Ryssens19,Gall94} and developing acceleration strategies based on the 
heavy-ball method~\cite{Ryssens19b}. The resulting iterative process is
much more robust than direct diagonalisation and sufficiently reliable for 
large-scale automated blocked calculations in odd-mass and odd-odd nuclei. 
Further details on this implementation will be presented elsewhere.

An important difference between the direct diagonalisation approach and the gradient 
method is the nature of the many-body state that can be targeted. The direct 
diagonalisation approach allows for the selection of one or more blocked 
quasiparticles at every iteration. Depending on the selection strategy and with 
a careful implementation, one can use the direct diagonalisation strategy to 
construct multiple auxiliary states with a given set of conserved quantum 
numbers. The gradient method, on the other hand, solves for the state with the 
lowest overall energy among all states that are not orthogonal to the starting 
point of the evolution~\cite{Robledo11}; this means that the gradient method can
only construct one state among all those sharing a complete set of quantum numbers. 
While this limits the gradient method's use in spectroscopic applications, it 
makes it the ideal tool for large-scale calculations of ground states:
a single calculation is guaranteed to find the state of lowest energy with 
given quantum number(s), while a direct diagonalisation approach offers no such 
guarantee.

\subsubsection{Searching for the ground states}
\label{sec:limitations}

The mean-field energy can be varied in a straightforward way in the variational 
space spanned by Bogoliubov states, but the collective correction energy 
can not. We adopt here the semivariational strategy of Ref.~\cite{Scamps21}: in
any single MOCCa calculation, we perform a consistent minimization of $E_{\rm HFB}$
and add the correction energy $E_{\rm corr}$ perturbatively. For any given 
nucleus, we perform a number of such calculations constrained to combinations
of both quadrupole moments $\beta_{20}$ and $\beta_{22}$ that scan the relevant part
of the energy surface. Among these calculations, we select the overall minimum of 
the \emph{total} energy as our final result for the ground state energy. 
Our search in quadrupole deformation was performed
with a resolution of  $ \Delta \beta_{20} = \Delta \beta_{22} = 0.005$ and was 
restricted to values of $\beta_{2} < 0.6$ and $\gamma \in [0,60]^{\circ}$
\footnote{Our definition of the quadrupole deformations ($\beta_{20}$,$\beta_{22}$), or
equivalently $(\beta, \gamma)$ can be found in Ref.~\cite{Scamps21}.}.
By scanning the energy surfaces we do not only find the minima of the
total energy including the corrections, but also ensure that we locate the 
global minimum when a nucleus exhibits shape coexistence. 

For odd-mass and odd-odd nuclei, this search is performed for quasiparticle 
excitations of both positive and negative parity, but only for quasiparticle
excitations with $z$-signature $\eta = +i$ to reduce the total computational 
effort. During the search for the optimal quadrupole deformation, we perform 
two (four) MOCCa calculations for odd-mass (odd-odd) nuclei. 
For odd-mass nuclei, this restriction does not impact the generality of 
our results: all many-body states for such nuclei come in pairs connected 
by time-reversal that have identical energy but angular momentum pointing 
in opposite direction. For odd-odd nuclei however, the relative orientation of the 
angular momenta of the odd neutron and proton matters since the four possible 
combinations of neutron and proton $z$-signature are only related pairwise by 
time-reversal, such that configurations with $\eta_n \eta_p = \pm 1$ will not 
have exactly the same energy. Experiment strongly favours the intrinsic 
spins of the odd neutron and proton to be parallel, as evidenced by the success 
of the Gallagher-Moskowski coupling rules for strongly defor\-med 
odd-odd nuclei~\cite{Gallagher58,Boisson78}. In rare-earth nuclei, the 
observed splitting can be on the order of 100-400 keV~\cite{Boisson78}, but 
we have opted to not search this degree of freedom for the first 
exploration presented here. In any case, it has been pointed 
out in Ref.~\cite{Robledo14} that the standard forms of both Skyrme and Gogny 
EDFs might be unsuited to correctly model this effect.

Another limitation of our strategy that is not immediately apparent is
our restriction to values of $\gamma \in [0^{\circ}, 60^{\circ}]$. For 
time-reversal conserving calculations, values of $\gamma$ outside this range
represent different orientations of the nucleus with respect to the Cartesian 
axes of the simulation volume. This is not the case for odd-mass or 
odd-odd nuclei: the imposed $z$-signature $\hat{R}_z$ and $y$-time-simplex 
$\check{S}^T_y$ symmetries imply that the energy of a nucleus with given 
shape is not entirely independent of its orientation in the simulation volume 
due to the angular momentum of the blocked quasiparticle(s). This orientation
effect can in in some cases also be of the order of 100 keV for odd-mass nuclei
\footnote{To the best of our knowledge, the effects of alispin rotations on
the masses and other properties of odd-odd nuclei have 
not been studied so far.} and can be discussed in terms of the 
so-called `alispin' of the quasiparticle(s)~\cite{Schunck10}. 

\subsection{Moments of inertia and cranking calculations}
\label{sec:mois}

In Sec.~\ref{sec:model}, we will discuss the properties of BSkG2 concerning 
the rotational motion of deformed nuclei in terms of several quantities. To 
clarify our terminology, we discuss here briefly the three different kinds of
 moment of inertia (MOI) we employ below, as well as the concept of 
 self-consistent cranking calculations.

The first and simplest MOI is the Belyaev MOI $\mathcal{I}^{\rm B}_{\mu}$~\cite{Belyaev61}. An 
expression for this quantity in terms of Bogoliubov quasiparticles can be 
derived through the application of simple first-order perturbation 
theory~\cite{RingSchuck,Scamps21}, meaning that this MOI can be obtained from
a mean-field calculation without significant additional effort. For this reason,
it is the Belyaev MOI we employ in the expressions for the rotational and 
vibrational correction.  

However, a simple perturbative calculation cannot capture the response of 
the mean-fields, both time-even and time-odd, to rotation\footnote{The Belyaev 
MOI is problematic for another, more technical reason: the simple perturbation 
theory argument cannot be applied without modification for odd-mass and odd-odd 
nuclei, see Appendix B in Ref.~\cite{Scamps21}. \label{footnote:MOI}}. 
The Thouless-Valatin MOI $\mathcal{I}^{\rm TV}_{\mu}$
does encode this information~\cite{Thouless62}, and is generally somewhat 
larger than the Belyaev one~\cite{Petrik18}. However, calculating $\mathcal{I}^{\rm TV}_{\mu}$ 
requires more effort: one approach is to extract it from the spurious 
rotational modes obtained in (Q)RPA calculations for deformed 
nuclei~\cite{Petrik18,Thouless62,Hinohara15}. Such calculations have only 
rarely been performed for large numbers of nuclei and, to the best of our 
knowledge, never for odd-mass or odd-odd nuclei. Instead, this effect is often 
accounted for by a simple multiplication of the Belyaev MOI by a factor 1.32~\cite{Libert99}.

One can also model the rotational properties of nuclei on a mean-field level 
with self-consistent cranking calculations~\cite{Hellemans12,Gall94,Thouless62}:
instead of the energy $E_{\rm tot}$, one variationally optimizes the Routhian $R$
\begin{align}
R \equiv E_{\rm tot} - \omega_z \langle \hat{J}_z \rangle\, ,
\label{eq:routhian}
\end{align}
where $\omega_z$ is a Lagrange multiplier and $\hat{J}_z$ is the angular momentum 
operator around the $z$-axis. For non-zero values of $\omega_z$, 
the last term in Eq.~\eqref{eq:routhian} can be interpreted as an external field
that explicitly breaks time-reversal invariance, such that the minimum of 
Eq.~\eqref{eq:routhian} will have a finite angular momentum 
and all time-odd densities will generally not vanish. For a deformed nucleus, 
this angular momentum can be identified with collective rotation around the 
z-axis at a frequency 
of $\hbar \omega_z$. By solving Eq.~\eqref{eq:routhian} for different values of 
$\omega_z$ or, equivalently, for different values of $\langle \hat{J}_z\rangle$,
one can compare calculations to experimental data on collective rotational bands.
For cranking calculations, we orient the nucleus such that the z-axis coincides 
with the intermediate axis of the nucleus, whose MOI is the relevant one for 
collective rotation of even-even nuclei at low spin.

In particular, the Thouless-Valatin MOI is equivalent to the kinematical 
moment of inertia $\mathcal{I}^{(1)}$ obtained from cranking calculations 
at infinitesimally small frequencies $\omega_z$:
\begin{align}
\mathcal{I}^{\rm TV}_{z} = \lim_{\omega_z \rightarrow 0} \mathcal{I}^{(1)}_z 
                         = \lim_{\omega_z \rightarrow 0} \frac{\langle \hat{J}_z \rangle}{ \omega_z} \, .
\label{eq:TV_cranking}
\end{align}
This way,\footnote{This expression has the significant 
advantage that it is applicable to odd-mass and odd-odd nuclei without
any caveats, as opposed to the expressions for the Belyaev MOI usually found
in the literature, see footnote~\ref{footnote:MOI}.} 
obtaining $\mathcal{I}^{\rm TV}_z$ for a given ground state 
requires just one (time-reversal breaking) mean-field calculation at 
some small value of $\omega_z$.

The Belyaev and Thouless-Valatin MOI characterize the moment of inertia 
of the ground state. Another quantity of interest that characterizes the 
variation of rotational properties along an entire rotational band is the 
dynamical moment of inertia $\mathcal{I}^{(2)}$~\cite{Bohr81,Dudek92}. 
This quantity is defined as the inverse of the derivative of the energy with respect to the
frequency, i.e.\ $\mathcal{I}^{(2)} \equiv ( d E/d \omega)^{-1}$~\cite{Hellemans12,Gall94}.
Experimental information on this quantity can be obtained through taking finite
differences of in-band $\gamma$-ray transitions~\cite{Singh02}.

\subsection{Construction of the model}

\subsubsection{Ingredients of the objective function}
\label{sec:ingredients}

Four properties of finite nuclei enter the objective function of the parameter 
adjustment; two of these concern ground states. The most important ingredient 
is the set of 2457 known masses of nuclei with $Z \geq 8$ tabulated 
in AME2020~\cite{Wan21}. As discussed already in Ref.~\cite{Scamps21} for 
BSkG1 and in the context of the BSk models (particularly Ref.~\cite{Goriely06}), 
fitting only the masses results in excessively large neutron pairing strengths
that result in unrealistic predictions for observables other than masses, 
such as pairing gaps, fission barriers and level densities~\cite{Goriely06}.
To control the neutron pairing strength, we also adjust the calculated 
$uv$-averaged pairing gaps $\langle \Delta \rangle_n$ \cite{Bender00b,Scamps21} 
to the known values of the five-point neutron mass gaps $\Delta^{(5)}_{n}$
\begin{align}
\nonumber
\Delta^{(5)}_n(N,Z) \equiv& - \frac{(-1)^{N}}{8} 
                    \bigg[ M(N+2,Z)  \nonumber \\
&   -4 M(N+1,Z)     +6 M(N,Z)  \nonumber \\
&   -4 M(N-1,Z)     +\phantom{4} M(N-2,Z)    \bigg] \, .
\label{eq:def_delta5}                    
\end{align}
Since all discontinuities in the systematics of binding energies such as shell 
effects contribute to the experimental gap~\cite{Bender00b,Duguet2001b}, 
we restrict our fit of pairing gaps to nuclei at least four neutrons away 
from a magic number. As we will discuss in more detail below, we consider
this to be a more satisfactory way to control the pairing strength than the 
inclusion of the rotational MOI of heavy nuclei that were included in the objective 
function of BSkG1. 

Aside from information on ground states, we include two properties of 
nuclear fission into the parameter adjustment: (i) empirical values for the 
primary and secondary fission barriers of twelve even-even actinide nuclei 
tabulated in the RIPL-3 database~\cite{Capote09} and (ii) the excitation energy 
of the fission isomer for seven even-even actinide nuclei from Ref.~\cite{Goriely07}. 
Both the RIPL-3 database and Ref.~\cite{Goriely07} contain data on more nuclei
than those we included in the fit; we restrict ourselves to nuclei with 
barriers modelled as double-humped in RIPL-3, i.e.\ $Z \geq 90$.
These nuclei generally have primary fission barriers below 10 MeV and are thus 
much more of interest to astrophysical applications than lighter nuclei.
We also exclude the Th isotopes, as we did not know \emph{a priori} whether
or not their calculated ground states would exhibit a static octupole 
deformation~\cite{Ryssens19b}, which could lead to inconsistencies between the
assumptions we made for the calculation of ground states and barriers, respectively.
Finally, we restrict ourselves to even-even nuclei for simplicity.\footnote{We
recall that only very few papers attempting EDF calculations of the fission barriers
of odd nuclei can be found in the literature at all, and none of them allows for as
general shapes as we do.}  In summary, we use primary and secondary 
barriers of $^{232-234-236-238}$U, $^{238-240-242-244}$Pu 
and $^{242-244-246-248}$Cm, as well as the isomer excitation energies of $^{236-238}$U, $^{238-240-244}$Pu and 
$^{242-244}$Cm.

Following the Brussels-Montr{\'e}al protocol~\cite{Goriely16}, we also include 
several properties of infinite nuclear matter.  We (i) fix the symmetry 
coefficient $J = 32$ MeV to ensure a moderately stiff neutron-matter equation of
state (EOS) to support neutron stars of moderate mass,
(ii) set the coefficient of the density-dependent term $\gamma = 0.3$ 
to obtain a reasonable incompressibility of charge-symmetric nuclear matter, 
$K_{\nu} \in [230,250]$ MeV~\cite{Chabanat97,Colo04} and (iii) enforced an isoscalar effective 
mass $M^*_s/M \approx 0.84$~\cite{Cao06,Zuo02}. We also include qualitatively 
the description of charge radii by optimizing the Fermi momentum $k_F$ to the
data in the compilation of Ref.~\cite{Angeli13}. 

In summary, the data employed for the parameter adjustment of BSkG2 is very 
similar to that used for BSkG1~\cite{Scamps21}, but not identical. We have
(i) updated the masses to those of AME2020, (ii)
replaced the rotational properties of heavy nuclei from the objective function
by the average pairing gaps, and (iii) added information on fission properties.

\subsubsection{Two-step optimization with neural networks}

The optimization of the objective function described above is an enormous computational 
challenge: twenty-five parameters need to be adjusted on thousands of data 
points. We employ again the machine learning technique we developed for the 
parameter adjustment of Ref.~\cite{Scamps21}:  we train individual neural
networks on a growing library of Skyrme-HFB calculations, such that they 
can propose candidate parameter sets at little to no computational cost. 
Using a committee of hundreds of such networks, we can avoid the bias of any
given network and explore the parameter space efficiently.

Here, we facilitate the learning of the individual networks in a few ways 
compared to Ref.~\cite{Scamps21}. First, we now provide the networks with the value of 
$N_{\rm m}(N)$ and  $N_{\rm m}(Z)$, where $N_{\rm m}(n)$ counts the number of
magic numbers smaller than $n$. Such an input helps the networks recognize and emulate 
the discontinuous changes in nuclear structure arising near magic numbers. 
Second, we provide the networks with all products out of any two of the 25 
model parameters, to highlight possible correlations between parameters. 
Finally, we do not train on the absolute
value of any observable, but rather on the difference of said observable with
respect to the BSkG1 value. This reduces the complexity 
of the training, as we expect the difference between models to be a smoother
function than the difference between either of the models and experiment.

The inclusion of fission data in the objective function of our optimization 
procedure is straightforward. Several aspects of the calculation of the fission 
properties that enter the objective function, however, are much more involved 
than the calculation of ground state properties: data on ground states are 
extracted from a single EDF calculation, whereas fission barrier heights are 
differences between multiple EDF calculations. Worse, to determine the 
latter one has to identify a fission path in a multi-dimensional energy 
surface that allows for more general shapes than our ground state calculations.
Aside from the inherent computational cost of performing such calculations 
for any given interaction, there is also the issue of reliably repeating such 
calculations with little to no human intervention for multiple nuclei and
candidate interactions. Finally, it is not \emph{a priori} clear if the neural
 networks can reliably learn fission and ground state properties at the same time.

We have opted to sidestep these issues by employing a two-stage fitting procedure,
similar to the one of Ref.~\cite{Goriely07}. In a first stage, we adjusted the 
parameters of the model without any reference to the fission data, producing an 
initial set of model parameters that optimizes the performance of the model on 
masses and pairing gaps while being subject to the infinite nuclear matter constraints.
As an intermediate step, we calculated the full potential energy surfaces of
the twelve selected actinide nuclei using this parameter set. 
In a second phase of the parameter optimization, we adjusted the nine 
parameters of the correction energy $E_{\rm corr}$ to the complete objective
function while freezing all other parameters. This second step 
was not more computationally expensive than the first, as the 
variation of the fission properties could be obtained by recalculating the 
collective correction from the values of $\langle J_{\mu}^2 \rangle$ and 
$\mathcal{I}^{\rm B}_{\mu}$ tabulated in our calculation of the surfaces.

Key to the success of this strategy is the observation that our fit protocol
already yields almost realistic surface properties in the first step; this
is illustrated below by the fact that the BSkG1 model provides a decent 
description of actinide fission barriers without them being included in its
objective function. For such a starting point, the fine-tuning of the 
correction energy in the second step suffices to systematically improve on our 
description of fission properties.
That such a two-step procedure works is far from being automatically the 
case; in fact, many widely used parameterisations of the Skyrme EDF overestimate
fission barriers by up to 10 MeV~\cite{Jodon16}, which cannot be corrected for 
by the fine-tuning a collective correction whose variation between ground state
and saddle points is typically of the order of a few MeV. One of the keys to finding 
reasonable agreement for fission barriers already without including them in 
the objective function is the inclusion of the two-body part 
of the centre-of-mass correction in the EDF~\cite{Bender00,daCosta22}, 
as we do here.

\section{The BSkG2 mass model}
\label{sec:model}

\begin{table}[t]
\caption{
The BSkG2 parameter set: sixteen parameters determining the self-consistent
mean-field energy $E_{\rm HFB}$ and nine determining the correction energy 
$E_{\rm corr}$. For comparison, we include the values of the BSkG1 parameter
set~\cite{Scamps21}. Note that instead of parameter $x_2$ we list the 
values of the product $x_2 \, t_2$.
}
\begin{tabular}{l|d{6.8}|d{6.8}|d{6.8}}
\hline
 \mc{Parameters}                       &  \mc{ BSkG1 }   &  \mc{ BSkG2 } \\
\hline
$t_0$         [MeV fm$^3$]             & -1882.36        & -1885.74      \\
$t_1$         [MeV fm$^5$]             &   344.79        &   343.59      \\
$t_2$         [MeV fm$^5$]             &    -2.43198     &    -8.04132   \\
$t_3$         [MeV fm$^{3 + 3\gamma}$] & 12322.0         & 12358.4       \\
$x_0$                                  &     0.196276    &     0.181775  \\
$x_1$                                  &    -0.580308    &    -0.584003  \\
$x_2 t_2$     [MeV fm$^5$]             &  -170.203       &  -162.003     \\
$x_3$                                  &     0.120751    &     0.101596 \\
$W_0$         [MeV fm$^5$]             &   123.922       &   108.655     \\
$W_0'$        [MeV fm$^5$]             &    83.519       &   108.603     \\
$\gamma$                               &     0.3         &     0.3       \\
\hline
$V_{\pi n}$   [MeV]                    &  -644.921       &  -483.366     \\
$V_{\pi p}$   [MeV]                    &  -682.559       &  -503.790     \\
$\eta$                                 &     0.692       &     0.486     \\
$\alpha$                               &     0.77        &     0.796     \\
$E_{\rm cut}$ [MeV]                    &     7.42        &     7.998     \\
\hline
$b$                                    &     0.930       &     0.878    \\
$c$                                    &     5.000       &     8.293    \\
$d$                                    &                 &     0.595    \\
$l$                                    &                 &     4.555    \\
$\beta_{\rm vib}$                      &                 &     0.788    \\
\hline
$V_W$         [MeV]                    &    -1.905       &    -1.805     \\
$\lambda$                              &   272.2         &   252.17      \\
$V_W'$        [MeV]                    &     0.671       &     0.745     \\
$A_0$                                  &    36.211       &    35.496     \\
\hline
\end{tabular}
\label{tab:param_skyrme}
\end{table}

\begin{table}[h]
\centering
\caption{
         Root-mean-square (rms) deviation $\sigma$ and mean deviation 
         $\bar \epsilon$ calculated from the sum over 
         $O_{\text{expt}} - O_{\text{model}}$
         for nuclear ground-state properties (first block) and fission 
         properties (second block) for the BSkG1 and BSkG2 models. These 
         values were calculated with respect to 2457 known masses $M$~\cite{Wan21} 
         of nuclei with $Z$, $N \geq 8$, 2309 neutron separation energies 
         $S_n$, 2173 $\beta$-decay energies $Q_\beta$, 884 measured charge radii 
         $R_c$ \cite{Angeli13}, 45 reference values for primary ($E_I$) and 
         secondary ($E_{II}$) fission barrier heights~\cite{Capote09} and 
         28 fission isomer excitation energies $E_{\rm iso}$ of actinide 
         nuclei~\cite{Goriely07}. 
         }
\begin{tabular}{l *{2}{d{6.5}}  }
\hline
\hline
Results                         &  \mc{ BSkG1 } &   \mc{ BSkG2 } \\ 
\hline
$\sigma_{\rm mod}(M)$ [MeV]     &   0.734       &  0.668  \\ 
$\sigma(M)$ [MeV]               &   0.741       &  0.678  \\ 
$\bar \epsilon (M)$ [MeV]       &  -0.026       & +0.026  \\ 
$\sigma(S_n)$ [MeV]             &   0.466       &  0.500  \\ 
$\bar \epsilon (S_n)$ [MeV]     &  +0.000       & -0.006  \\ 
$\sigma(Q_\beta)$ [MeV]         &   0.645       &  0.619  \\ 
$\bar \epsilon (Q_\beta)$ [MeV] &  +0.000       & -0.019  \\ 
$\sigma(R_c)$ [fm]              &   0.0239      &  0.0274 \\ 
 $\bar \epsilon (R_c)$ [fm]     &  -0.0008      & -0.007  \\ 
\hline
$\sigma(E_{\rm I })$ [MeV]           &  0.88 &  0.44 \\
$\bar{\epsilon}(E_{\rm I})$ [MeV]    & +0.80 & +0.24 \\
$\sigma(E_{\rm II})$ [MeV]           &  0.87 &  0.47 \\
$\bar{\epsilon}(E_{\rm II})$ [MeV]   & +0.71 & +0.10 \\ 
$\sigma(E_{\rm iso})$ [MeV]          &  1.00 &  0.49 \\
$\bar{\epsilon}(E_{\rm iso})$ [MeV]  & +0.67 & -0.36 \\
\hline
\hline
\end{tabular}
\label{tab:rms}
\end{table}

\subsection{Parameter values and global performance}

Table \ref{tab:param_skyrme} presents the values of all 25 parameters of the 
mass model, where the first group parametrizes the Skyrme energy, the 
second the pairing energy, the third the collective corrections and the last the 
Wigner energy. For comparison, we also give the corresponding BSkG1 parameters~\cite{Scamps21}. 
Note that the BSkG2 values of $W_0'$ and $W_0$ are almost identical, meaning that 
the parameter adjustment does not significantly exploit the liberty offered by the 
extended form of the spin-orbit EDF.

We show in Table~\ref{tab:rms} the mean ($\bar{\epsilon}$) and 
rms ($\sigma$) deviations of the BSkG2 model for all available 
experimental data for masses, neutron separation energies $S_n$, $\beta$-decay 
energies $Q_{\beta}$ and charge radii $R_c$. The performance of BSkG2 is 
essentially comparable to that of the BSkG1 model for all quantities, although 
the rms error on the absolute masses has improved, from 734 keV to 
678 keV. The inclusion of the time-odd terms in BSkG2 is not solely 
responsible for this modest improvement, since the inclusion of the 
vibrational correction introduces more parameters compared to BSkG1.
We note that the accuracy of both models for absolute binding energies 
is not as good as the one achieved by the later models in the BSk-family; for 
example the BSk27 model achieves an rms error $\sigma(M) = 0.517$ MeV on the 
AME2020 masses~\cite{Goriely13a}. We do not reach a similar accuracy 
here, mainly because we impose the symmetry coefficient $J=32$ MeV. For
 Skyrme EDFs of the standard form, choosing a lower value of $J$ around 
30~MeV tends to improve the overall systematics of masses~\cite{Goriely13b,Kortelainen10},
but results in an EOS for neutron matter that is incompatible with the most 
massive known neutron stars. The later BSk models\footnote{As an exception, 
the BSk27 model is based on a Skyrme EDF of standard form~\cite{Goriely13a}).} 
employ an extended form of the EDF that allows for a reconciliation of both 
constraints; we choose here the compromise value $J=32$ MeV~\cite{Chamel09}. 
The BSk models also differ from the approach here in the use of different 
pairing strengths for even and odd systems. Despite the difference in accuracy for 
the total masses, BSkG2 achieves a description of neutron separation energies 
and $\beta$-decay energies that is only slightly worse than that of BSk27, 
with $\sigma(S_n) = 0.429$ MeV and $\sigma(Q_{\beta}) = 0.524$ MeV
for the AME2020 dataset.

To verify the fission properties of the model, we extended our calculation of
fission barriers and isomer excitation energies from the twelve nuclei included
in the parameter adjustment to the complete set of forty-five
$Z \geq 90$ nuclei for which the RIPL-3 database lists reference values. 
The mean and rms errors for
these fission quantities are given in the second part of Table~\ref{tab:rms}.
Even though no information on fission properties entered its parameter adjustment, 
the BSkG1 model describes fission barriers in this region rather well, 
with rms errors on the primary and secondary barriers below 1 MeV. The model
is perhaps less suited to provide data for applications and extrapolation to 
neutron-rich nuclei, as it systematically underestimates both primary and
secondary barriers with mean deviations for both quantities on the 
order of 0.8 MeV. BSkG2 on the other hand achieves rms errors below 0.5 
MeV for primary and secondary barriers as well as isomers, while drastically 
reducing the mean deviations. This accuracy with respect to the RIPL-3
reference values is to the best of our knowledge unprecedented: other 
large-scale models achieve at best an rms deviation of about 0.6 MeV
on the primary barriers for the same set of nuclei, with generally larger
deviations between 0.7 and 1~MeV for secondary barriers and
isomers~\cite{Moller16,Goriely07,Mamdouh01,Giuliani18}.
A more detailed analysis and discussion of the fission properties 
of BSkG2 will be presented in a forthcoming paper~\cite{Ryssens22}.

\subsection{Nuclear masses}
\label{sec:binding_energies}

\begin{figure}
\centering
\includegraphics[width=.4\textwidth]{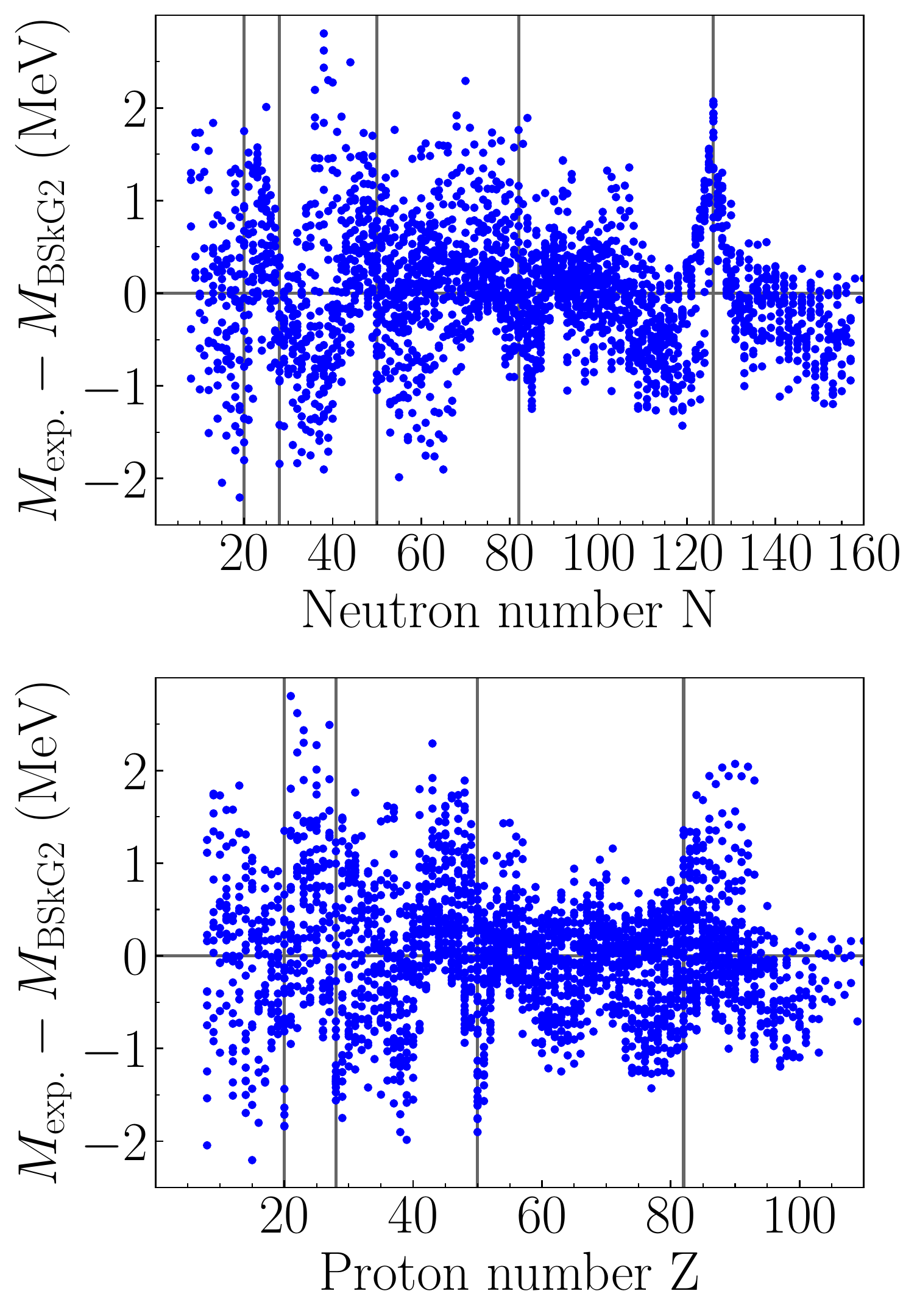}
\caption{(Color online) Difference of masses calculated with BSkG2 and 
          known values of Ref.~\cite{Wan21}, as a function of neutron number (top
         panel) and as a function of proton number (bottom panel). Gray bands 
         indicate the magic numbers.
         }
\label{fig:masses}
\end{figure}

We show the difference between experimentally known masses and the calculated
masses in Fig.~\ref{fig:masses} as a function of neutron number (top panel)
and proton number (bottom panel). Globally, the BSkG2 model achieves a good fit
to the data with only a handful of nuclei exhibiting a deviation that is larger
than 2 MeV. The largest deviations concern either light nuclei or nuclei close 
to the magic numbers; these patterns are similar to those of BSkG1 and the 
BSk-family of models~\cite{Goriely16}. 

\begin{figure}
\centering
\includegraphics[width=.4\textwidth]{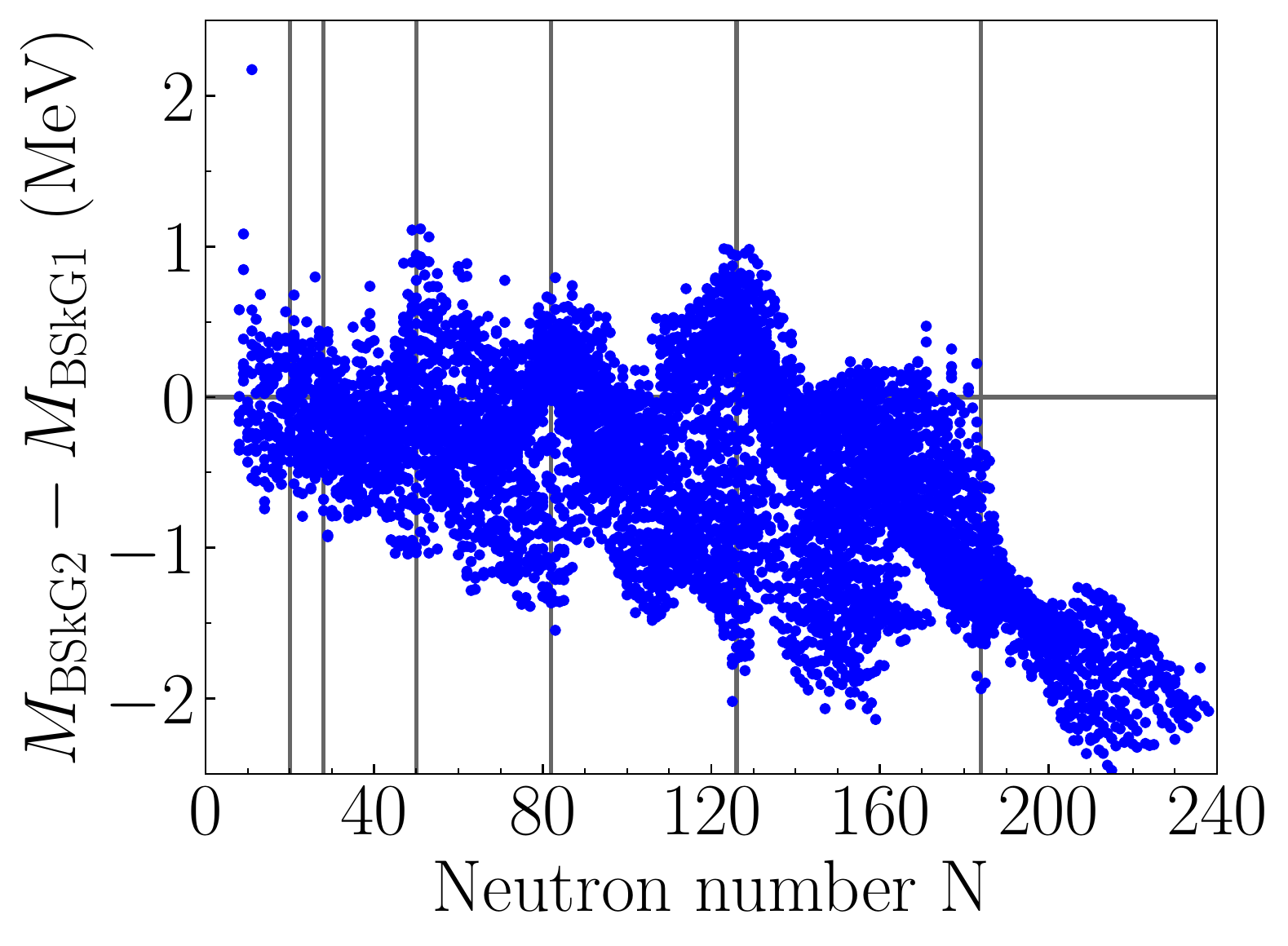}
\caption{(Color online) Difference between masses obtained with BSkG1 and 
          BSkG2 as function of neutron number. All nuclei $8\leq Z \leq 110 $ 
         that lie between the drip lines of
         both interactions are included. Gray vertical lines indicate the 
         neutron magic numbers, including $N=184$.}
\label{fig:diff_masses}
\end{figure}

The difference between the masses obtained with BSkG1 and BSkG2 for all nuclei 
within the drip lines for $8 \leq Z \leq 110$ are shown in Fig.~\ref{fig:diff_masses} as a function of 
neutron number. The newer model produces binding energies that are, on average, 
larger than the ones obtained from the BSkG1 model. This is reflected in the 
total number of nuclei: BSkG1 predicts the existence of 
6573 nuclei between proton- and neutron-drip line with $Z \leq 110$ while BSkG2 
predicts slightly more, 6719. The difference between both models exceeds two 
MeV systematically only for very heavy systems, beyond $N=184$.

\begin{figure}
\centering
\includegraphics[width=.45\textwidth]{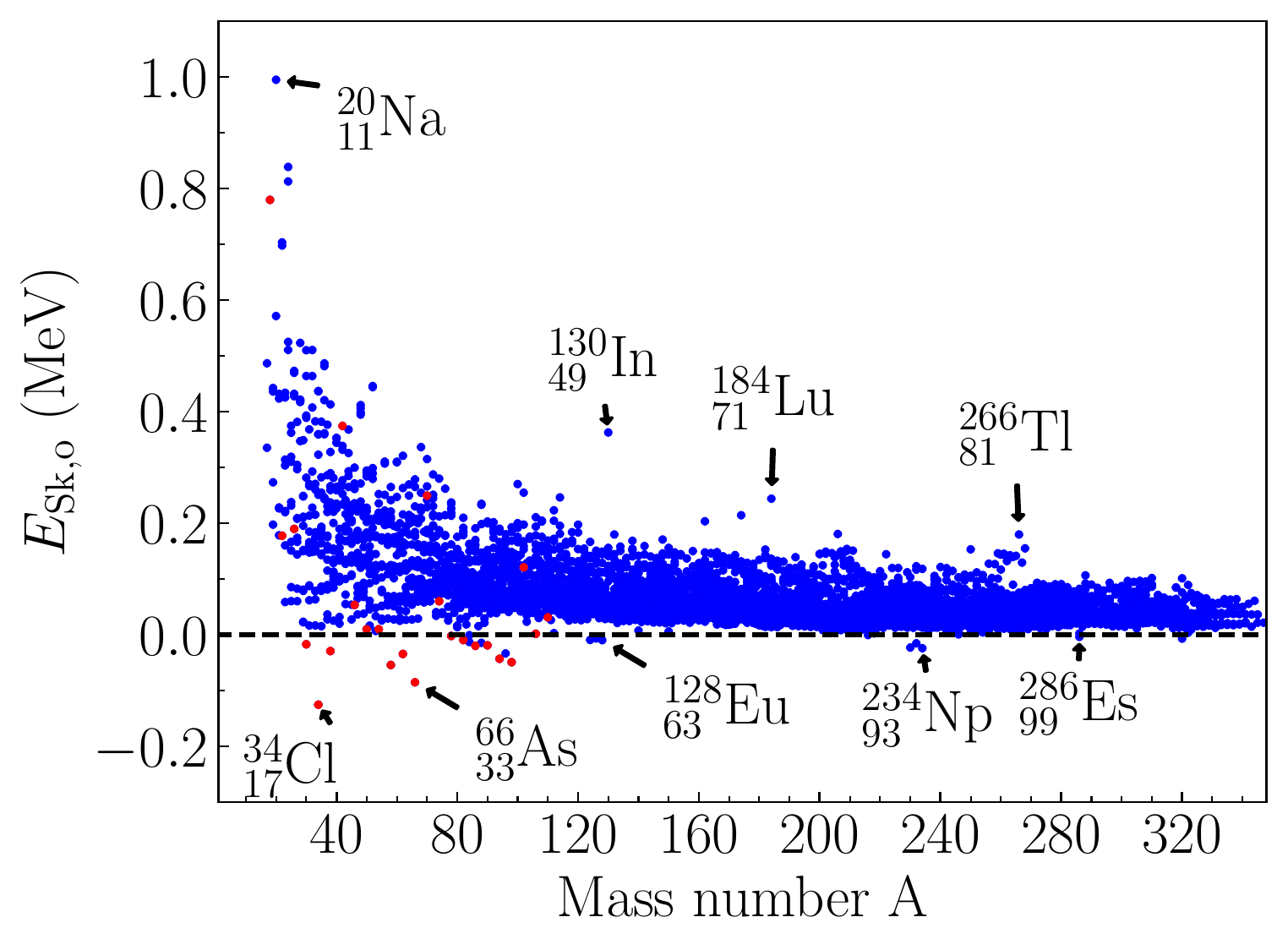}
\caption{(Color online) Time-odd energy $E_{\rm Sk, o}$ as a function of mass number $A$ for 
         all nuclei with odd neutron and/or proton number within the 
         calculated drip lines.
         $N=Z$ nuclei are indicated in red, all others in blue. Some nuclei
         with particularly large or small time-odd energies are pointed out. }
\label{fig:timeoddE}
\end{figure}

The contributions of the time-odd terms to the nuclear masses are shown in 
Fig.~\ref{fig:timeoddE}. Globally, the effects are small, although a few outliers can
be seen. The largest contributions occur for light nuclei, with a global 
maximum of about 1 MeV for odd-odd $^{18}$Na. For heavy nuclei, the largest 
effects are on the order of 200 keV. Nevertheless, in all regions of the nuclear chart there
are many nuclei for which the time-odd contribution to the energy is very small. 
For odd-odd nuclei, the average contribution of time-odd terms is about 
100 keV while that for odd-mass nuclei is slightly more than half that, roughly 
60 keV. The time-odd energy in even-$Z$, odd-$N$ nuclei is only slightly
larger than that in odd-$Z$, even-$N$ nuclei with on average 65 keV and 50 keV, 
respectively. Averaged over all nuclei affected, odd-mass and 
odd-odd, the time-odd terms reduce the binding energy by about 70 keV. These 
averages hide significant variations, as can be seen in Fig.~\ref{fig:timeoddE}.

While the total time-odd energy is never very large, we note that 
the contributions of individual terms in the energy density need not be so 
small themselves, particularly for light nuclei. For $^{34}$Cl for example, the 
total time-odd energy of about $-0.13$ MeV results from a cancellation between 
the contribution of spin terms (about +0.34 MeV) and the time-odd spin-orbit term 
(about $-0.42$ MeV) that are both a few times larger than the net effect. The 
current terms contribute very little for that nucleus.

A particularly striking aspect of Fig.~\ref{fig:timeoddE} is that the total 
time-odd energy is almost always positive, i.e.\ it reduces the total binding 
energy in virtually all nuclei that it contributes to. This can be easily 
understood in terms of the coupling constants of these terms that are listed
in Table~\ref{tab:todd:cpl}. In both isospin channels, the coupling constant 
$C^{s s}_t$ of the density-independent term is large and positive, while the
coupling constant of the density-dependent term $C^{s s \rho^{\gamma}}_t$ is
of similar size but opposite sign. The net effect is that for all densities
encountered in nuclei the spin terms are repulsive.
On the one hand, this is consistent with
empirical knowledge about collective spin- and spin-isospin response of 
nuclei, see Ref.~\cite{Bender02} and references therein, but on the 
other hand it is also known that specific proton-neutron matrix elements of 
the spin terms have to be attractive in order to reproduce the empirical
Gallagher-Moszkowski rule \cite{Robledo14}.
The coupling constants $C^{j \nabla s}_t$ of the time-odd spin-orbit 
contributions are both negative, but they multiply an integral that can have
either sign depending on the relative orientations of spin and orbital
angular momenta of the nucleons. Similarly, the contribution of the current 
terms can be of either sign. It often will remain small, however. For odd-mass 
nuclei, the current terms are dominated by the contribution from the 
blocked particle, for which the strictly positive integral over 
$\bold{j}^2_q(\bold{r})$ is multiplied by the very small sum of coupling constants 
$C^{j j}_0 + C^{j j}_1 = -4.264 \, \text{MeV} \, \text{fm}^{3}$. 
For odd-odd nuclei, the proton-neutron contribution will dominate as it is 
multiplied by the comparatively large difference
$2 \, (C^{j j}_0 - C^{j j}_1) = -38.554 \, \text{MeV} \, \text{fm}^{3}$
of isoscalar and isovector coupling constants, but depending on the relative
size and direction of the orbital angular momenta of the two blocked particles 
it multiplies an integral that can have either sign, such that this term is
also not necessarily attractive. In any event, the coupling constants
of the current terms remain smaller than those of the spin terms.

\begin{table}[t!]
\centering
\caption{
         Values of the isoscalar ($t=0$) and isovector ($t=1$) coupling 
         constants of the time-odd terms in the Skyrme EDF~\eqref{eq:Skyrme_to} found for BSkG2. The 
         $C^{s s}_t[n_0]$ are the effective spin-spin coupling constants 
         at saturation density $n_0$, see
         \ref{app:couplingconstants}.
         }
\begin{tabular}{l *{2}{d{7.3}}  }
\hline
\hline
Coupling constant   & $t=0$ &   $t=1$ \\ 
\hline
$C^{s s}_t$               [MeV fm$^{3}$]         &  150.023     &   235.718 \\
$C^{s s \rho^{\gamma}}_t$ [MeV fm$^{3+3\gamma}$] & -205.152     &  -257.468 \\
$C^{s s}_t[n_0]$          [MeV fm$^{3}$]         &   32.153     &    87.791 \\
$C^{j j}_t$               [MeV fm$^{5}$]         & -21.409      &    17.145 \\
$C^{j \nabla s}_t$        [MeV fm$^{5}$]         & -81.478      &   -27.151 \\
\hline
\hline
\end{tabular}
\label{tab:todd:cpl}
\end{table}

Further trends can be found by more detailed inspection of Fig.~\ref{fig:timeoddE}.
For instance, the effects of the time-odd terms are in general larger for nuclei
with odd proton and odd neutron number than for odd-mass nuclei. Furthermore, 
the outliers with largest positive $E_{\rm Sk., o}$ are odd-odd nuclei with 
both $N$ and $Z$ close to magic numbers. $^{130}$In and $^{184}$Lu are remarkable:
in both cases the isovector current term is comparably large (roughly $100$ and $60$
keV respectively), due to the the large angular momenta of the odd proton and neutron
oriented in opposite directions.

The size of the time-odd contribution to the nuclear binding energy and its trend 
with mass number is consistent with the existing literature on Skyrme parameterizations. 
The results of those that have been studied earlier, however, do not agree among
each other. For some, the effect has been reported as being
generally attractive, while for others it is generally repulsive or of varying 
sign. The comparison is also complicated by different groups having different 
strategies to choose the coupling constants of the spin terms that are not fixed 
by Galilean invariance \cite{Hellemans12,Bender02,Pototzky10,Schunck10}.
By contrast, the size of coupling constants of 
the current terms is determined by the effective mass of the parameterization, 
as these terms are linked to the time-even $\rho_t \tau_t$-terms 
of Eq.~\eqref{eq:Skyrme_te} by Galilean invariance. 
For example, $C^{jj}_0$ will vanish for a parameterization with isoscalar 
effective mass $M^*_s/M = 1.0$, and become increasingly negative when 
lowering $M^*_s/M$.

Our results are comparable to those of the large-scale HF+BCS study
of odd nuclei with three Skyrme parameterizations reported in Ref.~\cite{Pototzky10}: in
particular the authors find that including the time-odd spin terms, as we do here,
generally leads to a decrease in the total binding energy. The authors of
Ref.~\cite{Schunck10} find time-odd energies of about 100 keV in systematic 
calculations of both ground-state and excited states of odd-mass nuclei in the 
rare earth region. For different parameterizations and different treatments of 
time-odd terms, this effect can be either repulsive or attractive. Results for smaller 
sets of nuclei paint a similar picture; the time-odd terms of
the popular SLy4 parameterization~\cite{Chabanat98} lead to a decrease of 
binding energies of Ce isotopes by about 100-200 keV~\cite{Duguet2001a}, but to 
a few hundred keV for light nuclei~\cite{Satula99}.

The overall situation is different for relativistic EDF models: for the ones that
have been used in time-reversal-breaking calculations of odd- and odd-odd nuclei
so far, the time-odd terms always increase the binding energy by about 
100-300 keV, almost independent of the parameterization employed~\cite{Afanasjev10a}. 
The main difference to the non-relativistic approach that we use here is that 
the EDFs used in Ref.~\cite{Afanasjev10a} are 
constructed as Hartree models that target the time-even terms, and for which only 
those time-odd terms appear that are necessary to conserve the Lorentz-invariance
of the EDF~\cite{Vretenar05}. As a consequence, the time-odd sector of these models
only contains current and spin-orbit terms, but no spin terms, as can be easily 
seen when constructing their non-relativistic limit~\cite{Sulaksono07}. 
In fact, in order to describe spin- and spin-isospin response within these 
models, one is obliged to add additional phenomenological spin 
terms~\cite{Paar04a} that are not considered when calculating ground states. 
As relativistic EDF models all have comparatively low effective 
mass~\cite{Vretenar05}, the isoscalar current terms fixed by Lorentz invariance
are large and attractive, which explains the global energy gain from time-odd 
terms. Nevertheless, the overall size of the
effect and its decrease with mass number compares well to our results.
We are not aware of any systematic study of the effect of time-odd terms for 
EDFs of the Gogny type.

\begin{figure}[t!]
\centering
\includegraphics[width=.48\textwidth]{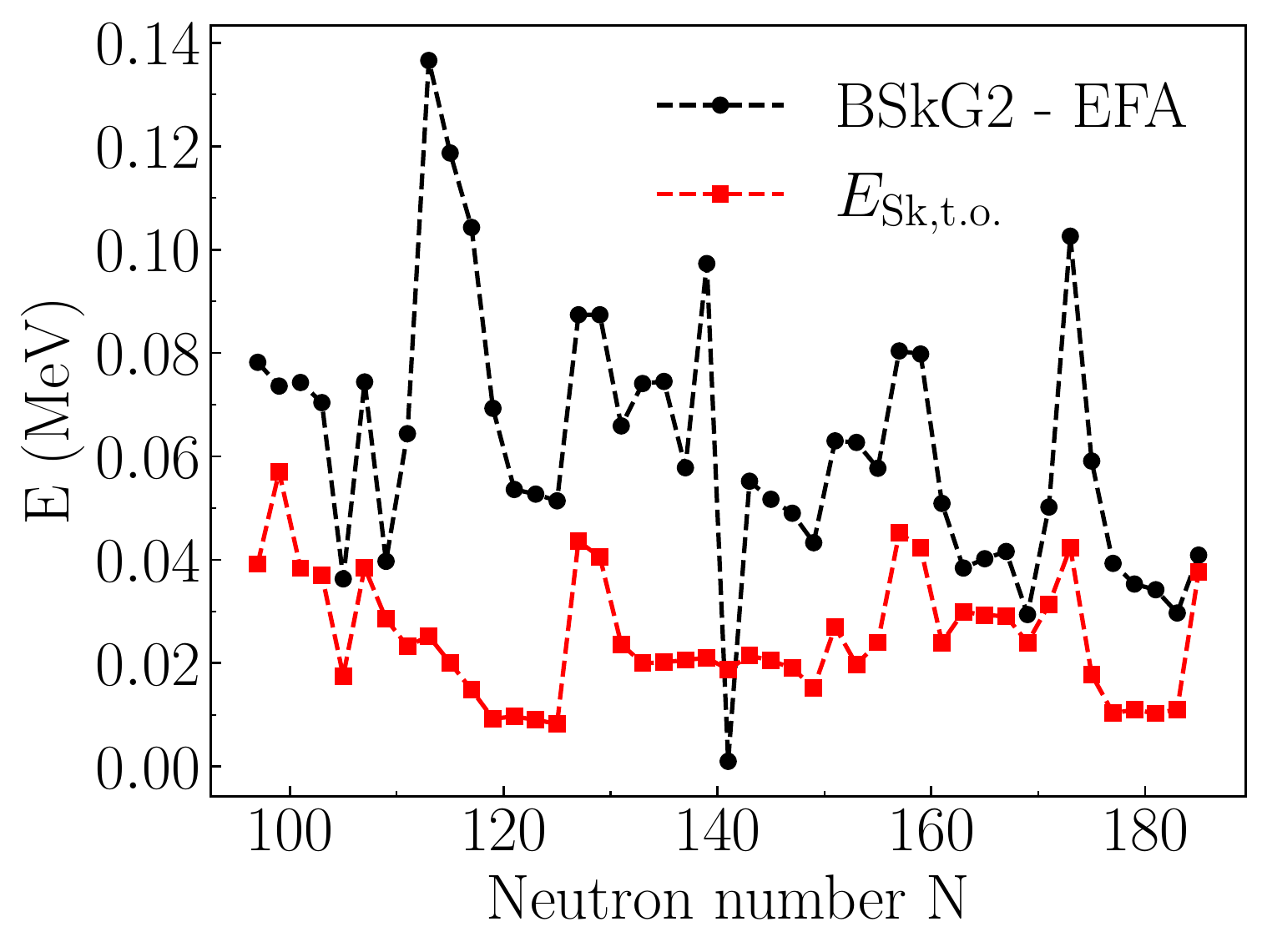}
\caption{ (Color online) Differences in $E_{\rm tot}$ as a function of neutron 
           number for all bound odd-mass Pb isotopes between the complete 
           calculation and a calculation employing the EFA (black circles).
           The total energy shift
           due to time-odd terms in the complete calculation (red squares) is 
           also shown.
        }
\label{fig:mass_diff}          
\end{figure}

\subsection{Quality of the equal filling approximation}
\label{sec:separation_energies}

We discuss in this section the quality of the EFA for absolute masses
and their differences. We will distinguish between two effects here:
(i) the direct contribution of the time-odd terms to the total binding 
energy and (ii) the polarisation effect, i.e.\ the change in the time-even 
part of the energy due to the presence of the time-odd mean-fields. 
We will visualize the direct effect by plotting the size of the time-odd terms 
and explore the second one by calculating mass differences between 
calculations employing the full EDF of Eq.~\eqref{eq:Eskyrme} and those 
employing the EFA. 
 
Figure~\ref{fig:mass_diff} shows the difference in $E_{\rm tot}$ for 
all bound odd-mass Pb isotopes between the BSkG2 drip lines as a function of 
neutron number, as well as the time-odd
contribution to the energy. First we note, as we already concluded from 
Fig.~\ref{fig:timeoddE}, that the EFA produces larger binding energies such 
that the difference plotted in Fig.~\ref{fig:mass_diff} is positive across
the entire isotopic chain\footnote{We note in passing that EFA calculations 
producing larger binding energies than complete calculations with blocking 
is not a violation of the variational principle. The EFA deals with statistical
mixtures of Bogoliubov states~\cite{Perez08}, while complete blocking 
calculations deal with pure Bogoliubov states that break time-reversal. Both 
types of calculations thus explore different variational spaces, neither of
which is a subspace of the other.
}.
 The time-odd terms show only a modest amount of
variation with neutron number and remain small everywhere. The 
complete effect can be several times larger than just the contribution of the 
time-odd terms, but it remains on the level of about a hundred keV. 
Compared to the rms deviation of the model on the total masses, 
$\sigma(M) = 0.668$ MeV, it seems unlikely the parameter adjustment was
significantly influenced by the inclusion of the time-odd terms.

\begin{figure}[t!]
\centering
\includegraphics[width=.48\textwidth]{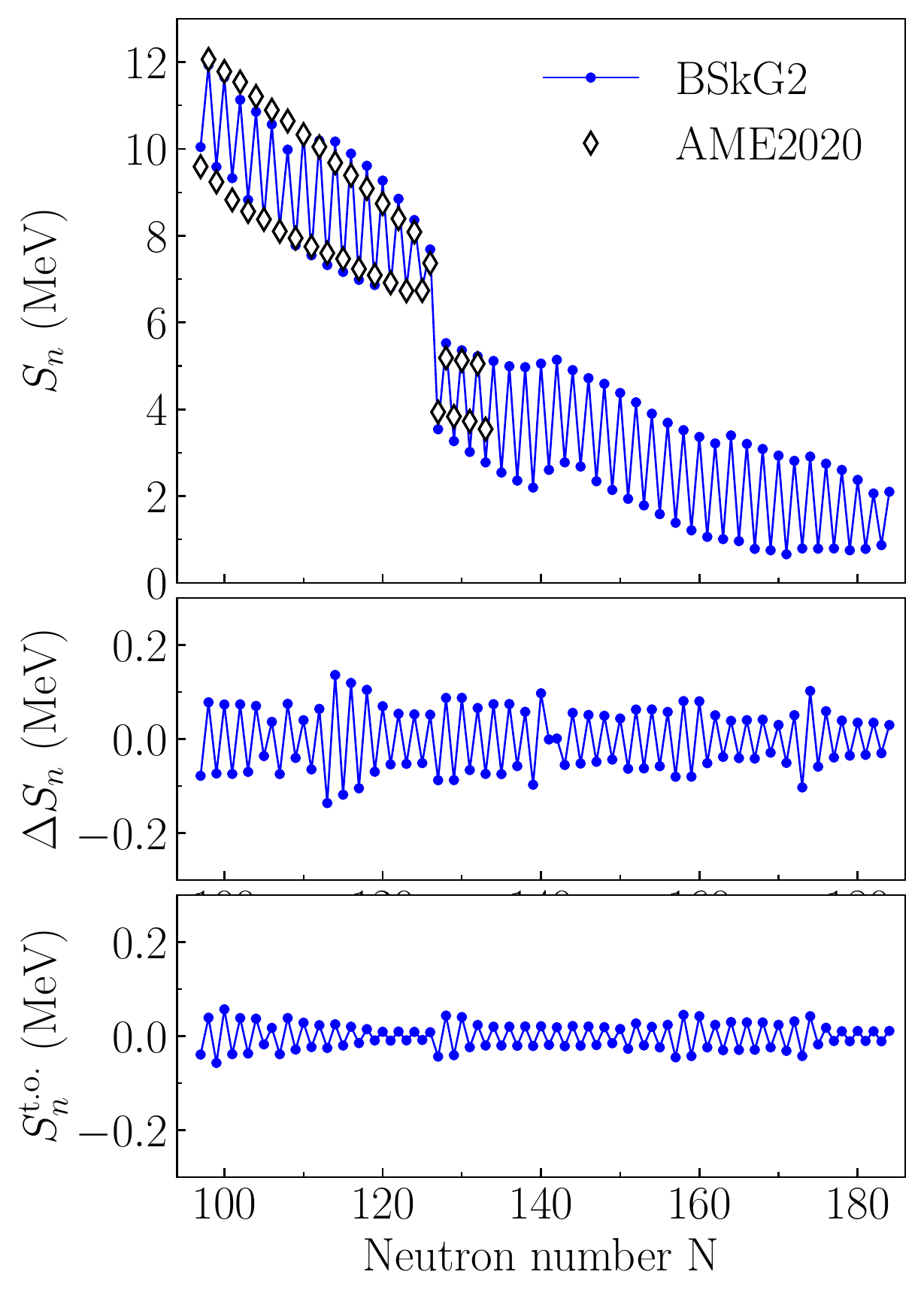}
\caption{ (Color online)
        Top panel: neutron separation energies $S_n$ along the Pb isotopic 
        chain as calculated with BSkG2 up to the last neutron-bound isotope 
        (blue circles) and as tabulated in 
        AME2020~\cite{Wan21} (open diamonds).
        Middle panel: difference in neutron separation energies $\Delta S_n$ between full and EFA calculations. 
        Bottom panel: contribution of the time-odd terms to the neutron separation 
        energy, Eq.~\eqref{eq:Snto}. }
\label{fig:separation_energies}          
\end{figure}

We show the calculated BSkG2 neutron separation energies for Pb isotopes in the 
top panel Fig.~\ref{fig:separation_energies}, as well as the data of AME2020.
As expected from the global accuracy of the model, the reproduction of the 
experimental separation energies is quite good for isotopes up to the $N=126$ 
shell closure, while for the heavier ones their staggering is somewhat 
overestimated. Calculation in EFA result in a curve that is indistinguishable
from the one shown on the top panel of Fig.~\ref{fig:separation_energies}. 
Instead, we show the difference between the complete and an EFA 
calculation, $\Delta S_n = S^{\rm BSkG2}_n - S^{\rm EFA}_n $, separately in the 
middle panel. The bottom panel of Fig.~\ref{fig:separation_energies} shows the
contribution of the time-odd terms to the neutron separation energies:
\begin{align}
S^{\rm t.o.}_{n}(Z,N) &= E_{\rm Sk,o}(Z,N-1) - E_{\rm Sk,o}(Z,N) \, . 
\label{eq:Snto}
\end{align}
As the time-odd terms are zero for even-even nuclei, for an even-even isotope this 
quantity equals the total time-odd Skyrme energy of its odd neighbor with one
neutron less, while for an odd-mass isotope this is minus the time-odd Skyrme 
energy of the nucleus itself. When plotting this quantity, two consecutive
values therefore have the same size, but opposite sign, which explains the symmetric 
pattern found in Fig.~\ref{fig:separation_energies}.
A few things are readily apparent: first, the differences between complete
and EFA calculations are small. Second, these differences 
always enhance the staggering between even and odd systems, i.e.\ they enlarge
the separation energy for even-$N$ nuclei and reduce it for odd-$N$ nuclei.
Third, the bare value of the time-odd energy is responsible for only roughly
half of this effect; it accounts for a few tens of keV only, although
its contribution is somewhat larger for lighter nuclei. The remaining part
of the complete difference between EFA and complete calculations is therefore 
due to the polarising effect induced by the time-odd mean-fields. 
Fourth, all of these effects remain roughly constant in size for all values 
of $N$, although their relative importance changes: near the drip line the 
separation energies become smaller and hence
the time-odd fields become comparatively more important.
The smallest separation energy in Fig.~\ref{fig:separation_energies} 
is about 660 keV for $^{253}$Pb, about 5\% of which is due to the time-odd 
terms. 

\begin{figure}[t!]
\centering
\includegraphics[width=.45\textwidth]{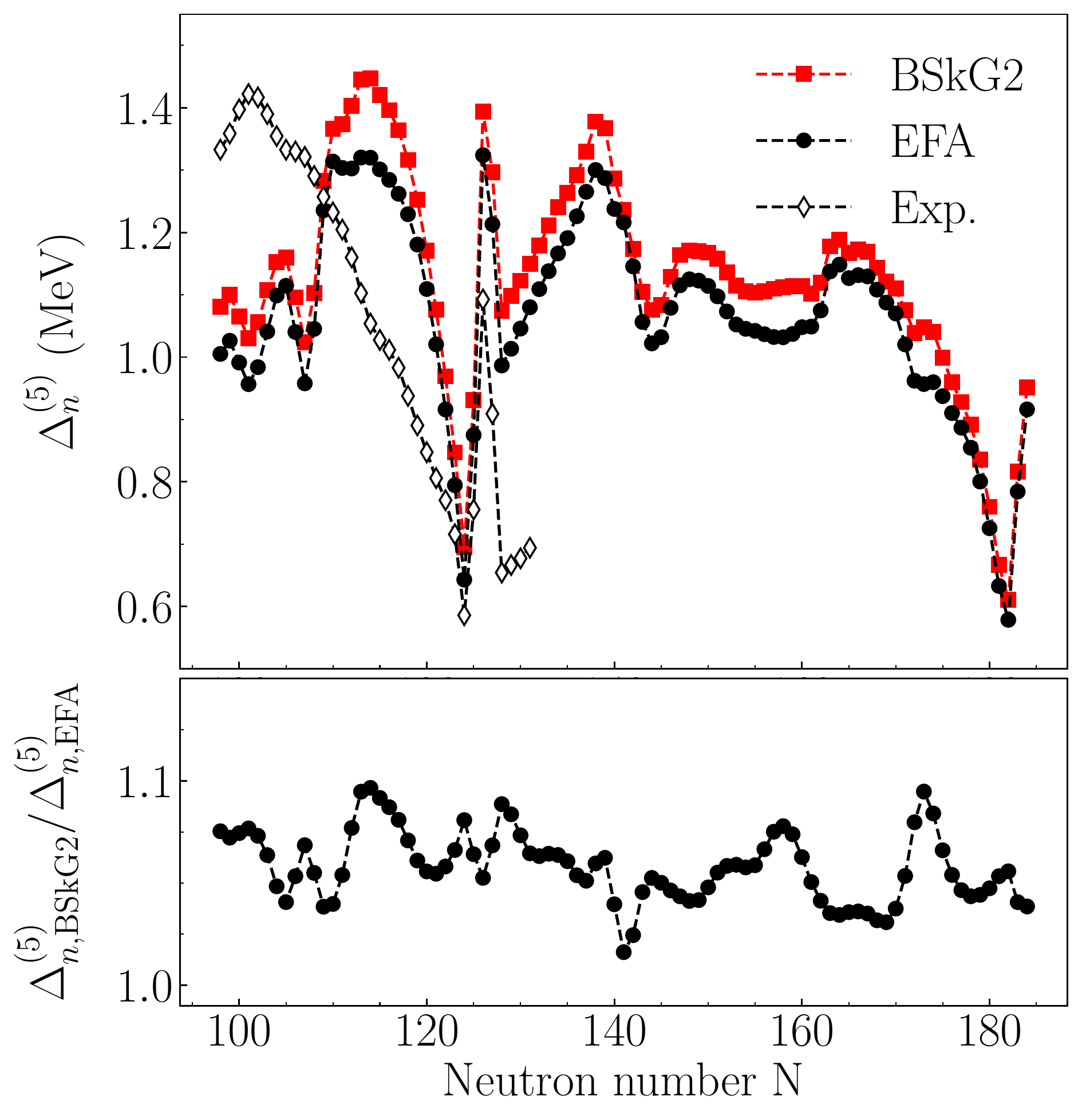}
\caption{ (Color online) 
        Top panel: neutron five-point pairing gaps $\Delta^{(5)}_n$ along the 
        Pb isotopic chain as calculated with BSkG2: the complete calculation
        (black circles) or when using the EFA (black circles). 
        Experimental data was taken from AME2020~\cite{Wan21} (open diamonds).
        Bottom panel: ratio of the $\Delta^{(5)}_n$ from the complete and 
        EFA calculations.
        }
\label{fig:delta5_Pb}          
\end{figure}

We show the effect of the inclusion of time-odd terms on the calculated 
five-point mass differences $\Delta^{(5)}_{n}$ of Eq.~\eqref{eq:def_delta5}
in the top panel of Fig.~\ref{fig:delta5_Pb} for the Pb isotopes. 
For the BSkG2 parameter set at least, accounting for time-reversal
symmetry breaking results an increase of the value of the five-point 
gap between 5\% and 10\% compared to an EFA calculation, as shown on the
bottom panel of Fig.~\ref{fig:delta5_Pb}. This increase remains constant 
from stability to the drip line. If one adjusts the pairing strength for a local study, 
i.e.\ limited to a handful of nuclei, an effect of this size can be meaningful. 
In the context of global calculations however, this difference is minor: a simple
rescaling by a few percent will not bring the global models closer to the 
experimental data in a systematic way, as can clearly from the qualitative 
rather than quantitative agreement between calculated and experimental values
for the neutron-deficient Pb isotopes in Fig.~\ref{fig:delta5_Pb}.

The authors of Ref.~\cite{Pototzky10} report an effect on the gaps of 
comparable size, though whether they increase or decrease 
depends on the details of the treatment of the time-odd terms. A much larger 
effect for a relativistic approach was reported in Ref.~\cite{Rutz99} for
spherical Sn isotopes. There is a qualitative difference between these
results and ours, though: As already mentioned above, the time-odd terms 
increases the binding energies of odd-mass nuclei in the relativistic approach
of Ref.~\cite{Rutz99}, such that at given pairing strength they decrease
the pairing gaps, while in our calculations they increase the gaps. Conversely, 
to obtain a given value for the pairing gap, in the relativistic approach 
of Ref.~\cite{Rutz99} one has to increase the pairing strength 
when including the effects of time-odd terms, whereas in ours the pairing strength 
has to be decreased. This can have a sizeable impact on many other observables.

A study employing a Gogny-type EDF found almost no effect on pairing gaps when 
including time-odd terms~\cite{Robledo12}. 
\footnote{Our comparison to these references is somewhat indirect: 
only Ref.~\cite{Rutz99} uses five-point gaps as we do, while 
Refs.~\cite{Pototzky10,Afanasjev10a,Robledo12} chose to analyse
three-point gaps instead.
}

\subsection{Pairing properties}
\label{sect:pairing:properties}

\begin{figure}
\centering
\includegraphics[width=.45\textwidth]{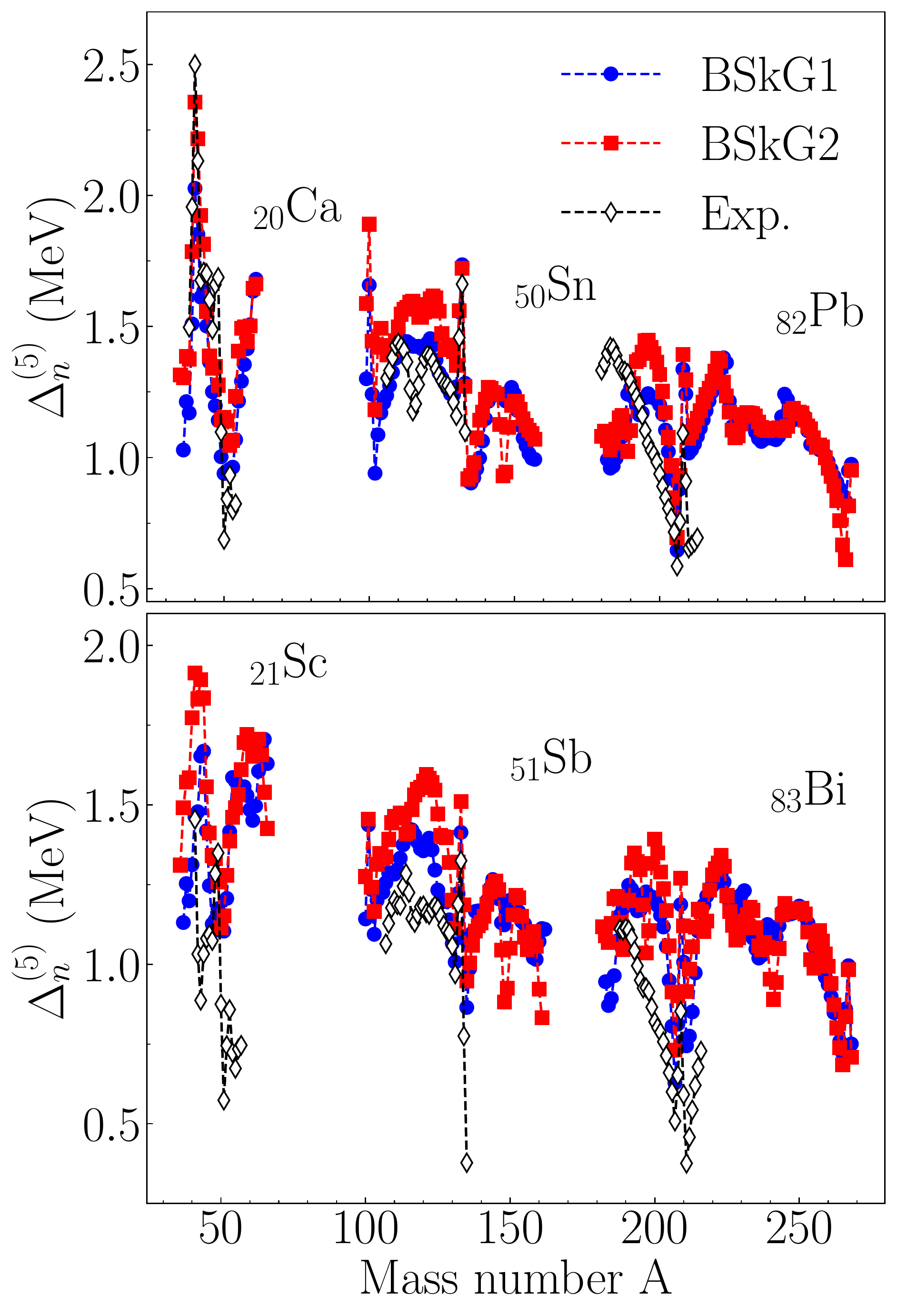} 
\caption{(Color online) Five-point neutron pairing gaps $\Delta_{n}^{(5)}$ for 
          selected even-$Z$ isotopic chains 
          (top panel: Ca, Sn, Pb) and neighbouring odd-$Z$ isotopic chains 
          (bottom panel: Sc, Sb and Bi) obtained with BSkG1 (blue circles), 
          BSkG2 (red squares) as well as experimental data (black triangles).
          }
\label{fig:delta5}     
\end{figure}

The top panel of Fig.~\ref{fig:delta5} shows the five-point neutron
gap $\Delta^{(5)}_n$ as defined in Eq.~\eqref{eq:def_delta5}
for the even-$Z$ Ca ($Z=20$), Sn ($Z=50$) and Pb ($Z=82$) 
chains (top panel) as well as the odd-$Z$ Sc ($Z=21$), Sb ($Z=51$) and Bi ($Z=83$)
chains (bottom panel). This quantity is generally indicative of the strength of 
pairing correlations in nuclei, but 
other effects such as time-odd terms (see the discussion around 
Fig.~\ref{fig:delta5_Pb}) and structural changes along the isotopic
chain contribute as well~\cite{Duguet2001a,Bender00b,Duguet2001b}. 
For the even-$Z$ nuclei, the overall size of the gaps is very reasonably
reproduced, but not all details of their evolution with neutron number.
For example, the calculated gaps increase much quicker than the experimental 
ones when going away from the $N=82$ and $N=126$ neutron shell closures in 
the Sn and Pb chains, respectively, which for the latter could 
also be clearly seen on Fig.~\ref{fig:delta5_Pb}. In addition,
the calculations miss some local features such as the dip observed near 
$A \simeq 115$ in the Sn isotopes and produce an arch-like structure around
$A \simeq 195$ in the Pb isotopes that is not present in the 
experimental data. Many of these local differences can be expected
to be related to imperfections in the description of the 
bunching of single-particle levels around the Fermi energy, rather
than to deficiencies of the modelling of pairing.
For neutron-rich Pb isotopes beyond $N=126$, the model appears to overestimate 
the neutron pairing gaps. The available data for the latter are limited however,
such that it cannot be entirely ruled out that this large difference is an artifact 
from the already mentioned too quick increase of the gaps around shell closures.
This quality of global reproduction is about typical for even-$Z$ nuclei across 
the nuclear chart and is entirely comparable to the performance of BSkG1. 
As already discussed in Ref.~\cite{Scamps21}, this quality of description of 
nuclear pairing can only be achieved 
by controlling it in the parameter adjustment; we did so here by fitting the 
calculated average pairing gap to five-point differences, as explained in 
Sec.~\ref{sec:ingredients}.

For odd-$Z$ chains, however, BSkG2 systematically overestimates the 
$\Delta^{(5)}_n$, as can be seen in the bottom panel of Fig.~\ref{fig:delta5} 
for the examples of the Sc, Sb and Bi isotopic chains. The global trends 
of the $\Delta^{(5)}_n$ are very similar to the ones found for the Ca, Sn, and Pb 
chains, but compared to experiment the values are systematically higher.
The same effect is also present for the proton gaps: experimental values 
are well-described for even-$N$ isotonic chains, but are somewhat too large 
for odd-$N$ chains. This deficiency is not a particularity of BSkG2: 
BSkG1 exhibits the same systematic effect. We do not interpret this 
mismatch between experiment and our models as a flaw of the adjustment of the 
pairing strengths but as the sign of a missing physical ingredient: in the
BSkG models, we miss a mechanism to produce extra binding energy due 
to the residual interaction between the odd neutron and odd proton in odd-odd 
nuclei~\cite{Jensen84,Friedman07,Wu16}. It is this additional binding energy 
which is thought to be at the origin of the observed systematic difference in 
three- or five-point gaps between both adjacent even-$Z$ and odd-$Z$ 
isotopic chains and adjacent even-$N$ and odd-$N$ isotonic chains. The 
microscopic-macroscopic approach of Ref.~\cite{Moller16} 
includes a simple analytic term to account for this effect: adding such a 
contribution here would shift the calculated curves on the bottom panel of 
Fig.~\ref{fig:delta5} downwards, improving our description of experiment.
For the older BSk-models, this effect was partially mimicked by adopting 
different pairing strengths for even-even, odd-even, even-odd and odd-odd 
nuclei~\cite{Goriely16}.

\begin{figure}[t!]
\includegraphics[width=.48\textwidth]{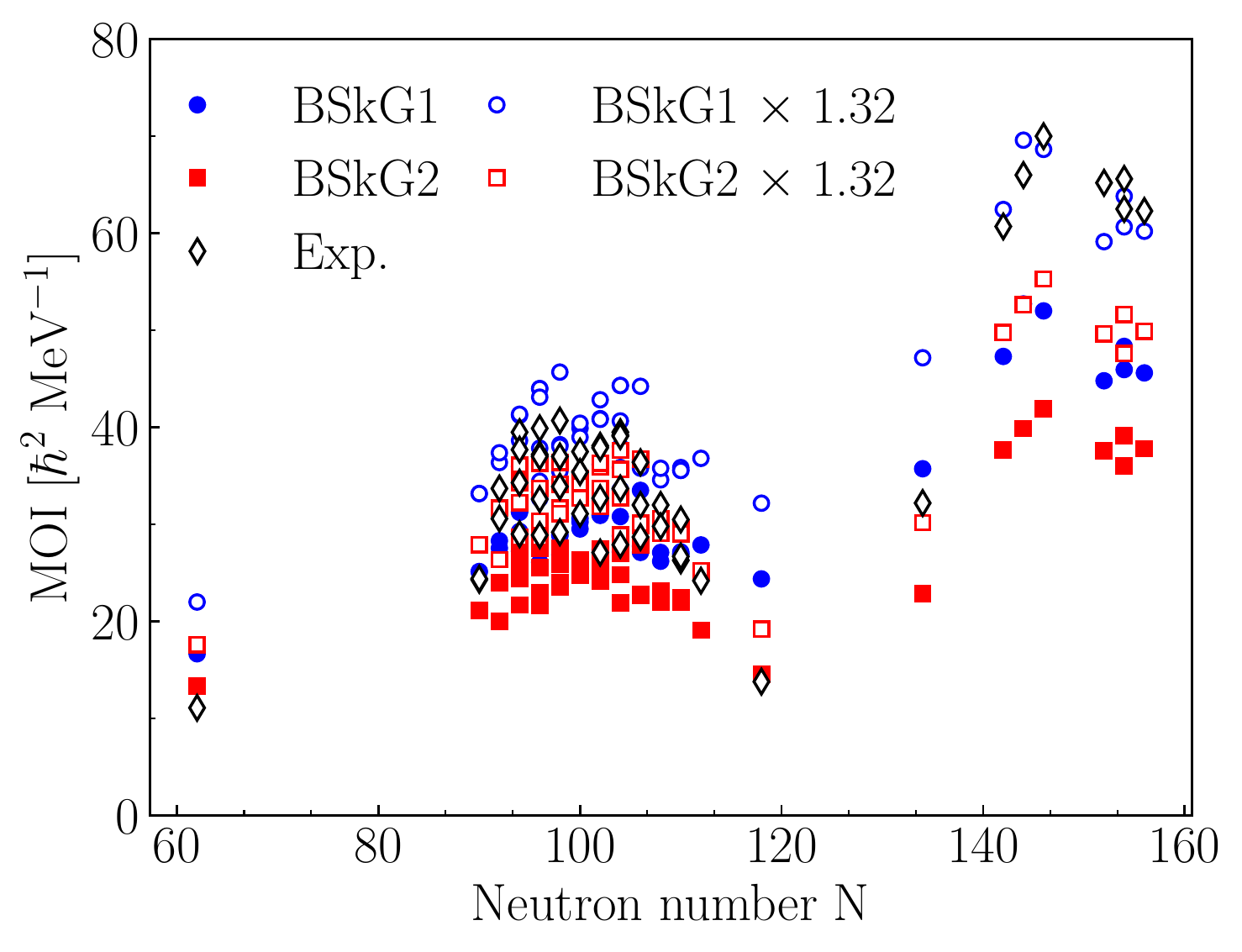}
\caption{Belyaev moments of inertia (MOI) as a function of neutron number 
          as obtained from BSkG1 and BSkG2 calculations (full blue circles and 
          red squares, respectively).  
          MOI multiplied with 1.32 are shown for both models as
          open symbols of the corresponding colors. 
          These results are compared to 
          experimental data for medium to heavy nuclei (open black diamonds) 
          from Refs.~\cite{Zeng94,Afanasjev00,Pearson91}. }
\label{fig:MOI}          
\end{figure}

Another quantity that is indicative of the pairing strength in nuclei are
the rotational moments of inertia (MOI): this quantity becomes 
systematically smaller with increasing pairing strength, leading to a 
larger spread between energy levels in a rotational band~\cite{Hellemans12,Belyaev61}.
The Belyaev MOI for 48 even-even nuclei were included in the BSkG1 objective 
function as a means to control the pairing strengths for protons and neutrons.
This approach is not entirely satisfactory: the Belyaev MOI 
captures only the perturbative first-order response to collective 
rotation and does not describe the entirety of the nucleus' response to rotation
as discussed in Sec.~\ref{sec:mois}. As outlined in Sect.~\ref{sec:mois}, the 
more reliable Thouless-Valatin approach to calculate the MOI systematically yields
values that typically are larger than the Belyaev MOI by about $30 \, \%$. 
Hence, a better way would be to include either a calculation of the Thouless-Valatin
MOI in the objective function (which is computationally much more demanding) or 
rescale the Belyaev MOI with an ad-hoc factor (which adds a phenomenological
ingredient to our model). Instead, we opted to drop the Belyaev MOI from the 
objective function, including the five-point gaps instead as (i) they are more 
directly connected to the neutron pairing strength and (ii) large amounts of 
experimental data is readily available.

In Fig.~\ref{fig:MOI}, we show the bare and rescaled Belyaev MOI for both BSkG1 and 
BSkG2. The bare BSkG1 values underestimate the experimental ones by about 
10\% for medium-heavy nuclei and an even larger percentage for actinide 
nuclei. Rescaling with the factor of 1.32 of Ref.~\cite{Libert99} fixes the 
discrepancy for the actinides, but renders the MOI for rare-earth nuclei 
somewhat too large by about 20\% on average.
The Belyaev MOI calculated with BSkG2 are systematically smaller than the
BSkG1 values. The bare values therefore significantly underestimate
the experimental ones for all nuclei across the nuclear chart. The rescaled 
MOI obtained with BSkG2, however, agree quite well with data for nuclei 
in the rare-earth region. Like with BSkG1, however, data for medium-heavy 
and actinide nuclei are not described simultaneously with BSkG2 either.

\begin{figure}[t!]
\includegraphics[width=.48\textwidth]{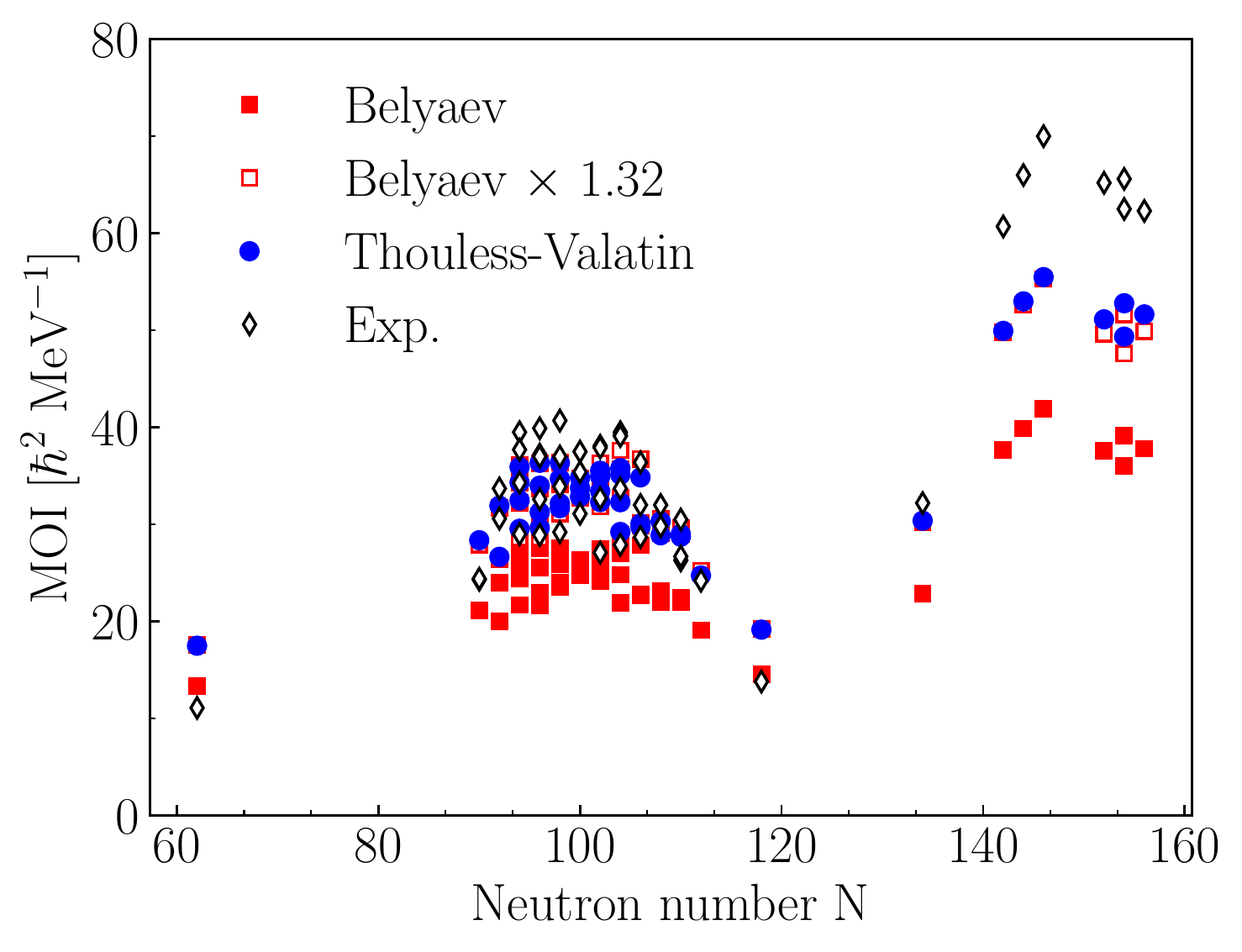}
\caption{
(Color online)
Thouless-Valatin (blue circles), Belyaev (full red squares) and rescaled 
Belyaev (open red squares) moments of inertia obtained with BSkG2 for the
nuclei already shown in Fig.~\ref{fig:MOI}.
Experimental data (open black diamonds) from Refs.~\cite{Zeng94,Afanasjev00,Pearson91}.
}
\label{fig:MOI_TV}
\end{figure}

Having fixed the time-odd terms in the parameter fit of BSkG2, we
also performed self-consistent cranking calculations minimizing the
Routhian of Eq.~\eqref{eq:routhian} at the small rotational frequency 
of $\hbar \omega_z = 0.01$ MeV for the same set of nuclei as shown 
in Fig.~\ref{fig:MOI}. For these calculations, we imposed an additional 
constraint to keep the quadrupole deformation of each nucleus fixed at 
its ground state value\footnote{Not keeping the deformation fixed results in 
only minor differences of typically less than 10\%.}. We show the 
Thouless-Valatin MOI obtained in this way, as well as the Belyaev and rescaled
Belyaev MOI in Fig.~\ref{fig:MOI_TV}. The agreement between experiment and 
Thouless-Valatin MOI is rather good, again with the exception of the actinides.
Altogether, these findings for the MOI indicate that the parameter adjustment 
of BSkG2 led to overall slightly smaller and thereby slightly more realistic 
effective pairing strengths than the ones of BSkG1, at least for not too heavy 
nuclei. We note though that the different values for 
$V_{\pi n}$ and $V_{\pi p}$ of BSkG1 and BSkG2 as listed in 
Tab.~\ref{tab:param_skyrme} are not directly indicative of the relative
pairing strength of the models, as also the ratio $\eta$ between volume and 
surface contributions to the pairing energy is quite different.

Figure~\ref{fig:MOI_TV} also confirms that rescaling the Belyaev MOI 
by a factor 1.32 is a good approximation to the Thouless-Valatin MOI for 
the well-deformed nuclei with $\beta \geq 0.26$ that we consider here.

\begin{figure}[t!]
\centering
\includegraphics[width=.4\textwidth]{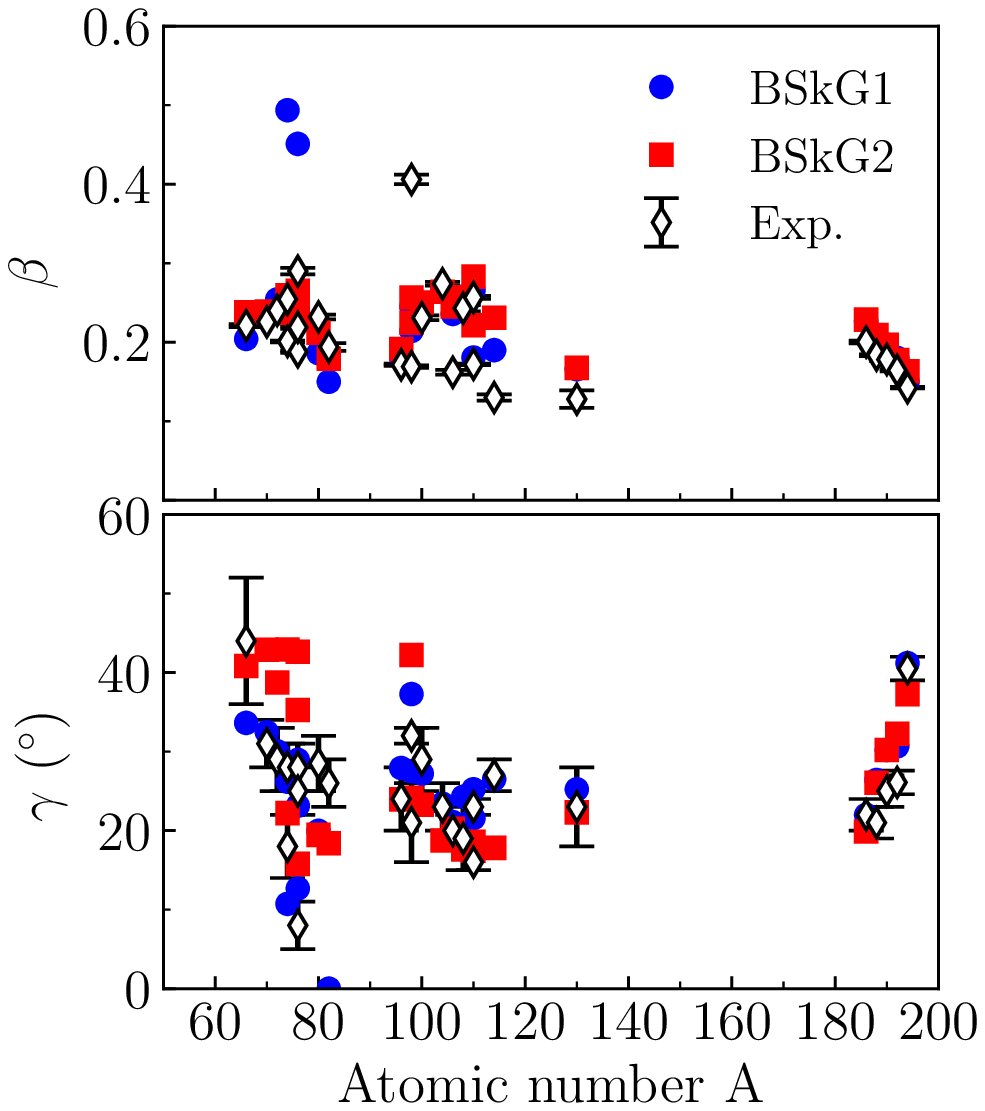}
\caption{ (Color online) Calculated quadrupole deformation $\beta$ (top panel) 
          and triaxiality angle $\gamma$ (bottom panel) with BSkG1 (blue circles)
          and BSkG2 (red squares), compared to experimental information 
          (open black diamonds).
          Top panel: quadrupole deformation $\beta$ with experimental
                        information from Nudat~\cite{nudat} for
                        $^{66}$Zn~\cite{Rocchini21},
                        $^{70}$Ge~\cite{Sugawara03},
                        $^{72}$Ge~\cite{Ayange16},
                        $^{74}$Ge~\cite{Toh00}, 
                        $^{76}$Ge~\cite{Ayange19},
                        $^{74,76}$Kr~\cite{Clement07},
                        $^{76,80,82}$Se~\cite{Kavka95},
                        $^{98}$Sr~\cite{Clement16},
                        $^{96}$Mo~\cite{ZielinskaPhD},
                        $^{98}$Mo~\cite{Zielinska02},
                        $^{100}$Mo~\cite{Wrzosek12},
                        $^{104}$Ru~\cite{Srebrny06},
                        $^{110}$Cd~\cite{Wrzosek20},
                        $^{114}$Cd~\cite{Fahlander88},
                        $^{106,108,110}$Pd~\cite{Svensson95},
                        $^{130}$Xe~\cite{Morrison20},
                        $^{186,188,190,192}$Os and $^{194}$Pt~\cite{Wu96}.
                        Bottom panel: triaxiality angle $\gamma$, with 
                        experimental data points extracted from measured sets of 
                        transitional and diagonal $E2$ matrix 
                        elements~\cite{Magda_priv} for the same nuclei.
          }
          \label{fig:triax_defo}
\end{figure}

\subsection{Shape: deformation and charge radii}

The shapes of nuclear ground states predicted by BSkG2 are very similar to 
the ones obtained with BSkG1. The new model reproduces about equally well the global 
systematics of (total) quadrupole deformation of even-even nuclei 
as deduced from measured $B(E2)$ transition rates~\cite{Raman01}. The systematics
of nuclear charge radii also do not differ strongly between
both models, as could have been expected from the comparable rms and mean 
deviations reported in Tab.~\ref{tab:rms}. We have also found the more 
fine-grained evolution of the radii with neutron or proton number in many 
regions to be similar.  Plots of deformation and charge radii 
compared to the experimental information on these quantities are  
close to Figs. 6 and 9 of Ref.~\cite{Scamps21}, we omit them here.

\begin{figure}[t!]
\centering
\includegraphics[width=.48\textwidth]{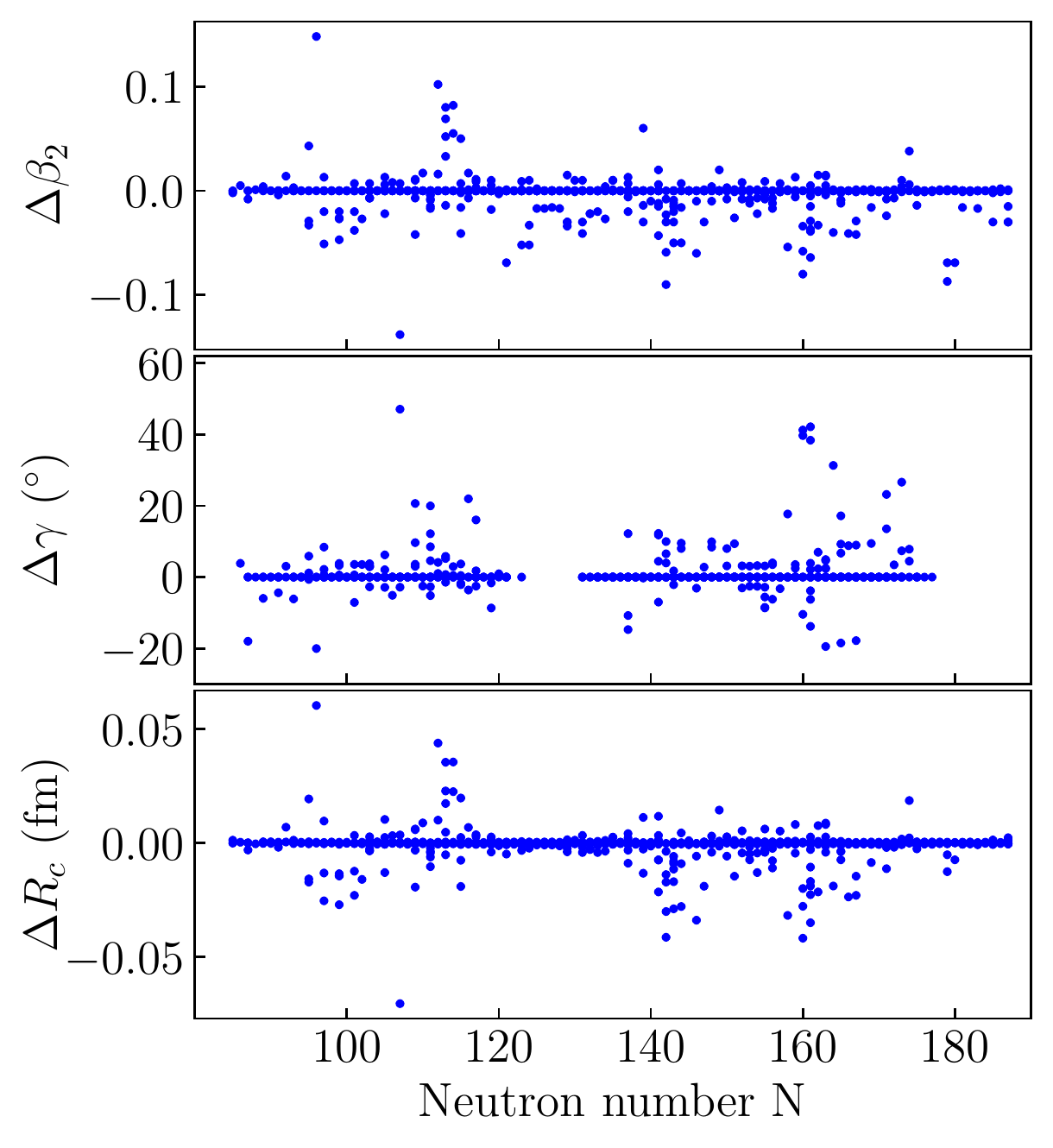}
\caption{ Difference in quadrupole deformation and rms charge radius between a 
          complete calculation and one employing the EFA (full - EFA) for odd-mass
          and odd-odd nuclei with $76 \leq Z \leq 88 $ between the BSkG2 drip lines. 
          Top panel: difference in $\beta$. Middle panel: difference in $\gamma$
          for nuclei with $\beta \geq 0.1$. Bottom panel: difference in 
          rms charge radius.}
\label{fig:polarization}
\end{figure}

This similarity between the models also extends to triaxial deformation: 
Fig.~\ref{fig:triax_defo} shows the total quadrupole deformation $\beta$ (top
panel) and the triaxiality angle $\gamma$ (bottom panel) for 26 even-even nuclei.
We compare (i) the calculated triaxiality angle with the mean value of 
$\gamma$ as obtained through the analysis of Coulomb excitation
experiments~\cite{Kumar72,Cline86,Magda_priv} and (ii) the calculated deformation 
$\beta$ with data from Nudat~\cite{nudat}.  For the heavier nuclei, and 
in particular the Os and Pt isotopes, both models yield almost identical results. 
For lighter nuclei, BSkG2 does not produce as large deformations for the two Kr isotopes, 
in better agreement with experiment than BSkG1. On the other hand, the $\gamma$ values 
calculated with BSkG2 for $A < 80$ compare less favorably with experiment than 
those of BSkG1. With the limitations of this type of comparison in 
mind~\cite{Scamps21}, we conclude 
that both models compare roughly equally well to the experimental data.

This discussion of the global aspects of deformation and radii makes one suspect
that the polarisation due to the time-odd terms does not strongly modify the 
nuclear shape. In order to quantify this effect, we show in 
Fig.~\ref{fig:polarization} the difference in nuclear shape due to the 
time-odd terms for 830 odd-mass and odd-odd nuclei with $76 \leq Z \leq 88$.
More precisely, we show the difference in total quadrupole deformation $\beta$
(top panel), triaxiality angle $\gamma$ (middle panel) and charge radius 
$R_c$ (bottom panel) between the complete calculation with time-odd terms 
included and another one employing the equal filling approximation. 

The change in total deformation due to polarisation is generally speaking 
limited, with $\Delta \beta$ exceeding $0.05$ only for about twenty nuclei\footnote{
Many nuclei are plotted at exactly zero change in $\beta$ and $\gamma$: 
our semivariational approach is limited by an accuracy of 
$\Delta \beta_{20} = \Delta \beta_{22} = 0.005$ as in Ref.~\cite{Scamps21}.}.
Only three outliers show $|\Delta \beta| \geq 0.1$: $^{177,187}$Tl and $^{195}$Bi.
All three lie in a region where multiple mean-field minima with different 
deformations but almost identical energies coexist; a small change in the conditions
of the calculations (i.e.\ presence versus absence of time-odd polarisation) can 
have a large impact on the nuclear shape by inverting the order of the minima. 
Experimentally, this region of neutron-deficient isotopes with $Z \simeq 82$ is known
for dramatic changes in charge radii with neutron number that are generally 
interpreted in terms of such closely-balanced minima~\cite{Sels19,Barzakh21}.
Changes in deformation in the top panel of Fig.~\ref{fig:polarization} 
correspond one-to-one to changes in charge radius, as shown in the bottom panel. 
Note that a change $\Delta R_c$ of about $0.05$ fm in this region of the nuclear
chart corresponds to a change in isotopic shift of about $0.25$ fm$^2$, which is
about the size of the anomalous odd-even staggering in the neutron-deficient Hg isotopes~\cite{Sels19}. 
Our conclusion is similar for the changes in triaxiality
angle $\gamma$: few nuclei exhibit a change larger than $10^{\circ}$ between a
complete calculation and one employing the EFA. Around $N=160$, a small clump 
of Tl and Bi isotopes stands out which change from nearly oblate to nearly 
prolate with $\Delta \gamma \approx 40^{\circ}$.

\begin{table}[t!]
\centering
\caption{Infinite nuclear matter properties for BSkG1~\cite{Scamps21} and BSkG2
         parameterizations. See Refs.~\cite{Goriely16,Margueron02,Chamel10} 
         for the various definitions.}
\begin{tabular}{l *{2}{d{6.5}} }
\hline
Properties             &  \mc{ BSkG1 } &      \mc{ BSkG2 }    \\
\hline
$k_F$~[fm]             &    1.3280     &    1.3265  \\
$n_0$~[fm$^{-3}$]      &    0.1582     &    0.1577  \\
$a_v$~[MeV]            &  -16.088      &  -16.070   \\
$J$~[MeV]              &   32.0        &   32.0     \\
$L$~[MeV]              &   51.7        &   53.0     \\
$M_s^*/M$              &    0.860      &    0.860   \\
$M_v^*/M$              &    0.769      &    0.773   \\
$K_v$~[MeV]            &  237.8        &  237.5       \\
$K_{\text{sym}}$~[MeV] & -156.4        & -150.6     \\
$K^\prime$~[MeV]       &  376.7        &  376.3     \\
$G_0$                  &    0.35       &    0.36    \\
$G_0^\prime$           &    0.98       &    0.98    \\
\hline
\end{tabular}
\label{tab:inm_prop}
\end{table}

\subsection{Infinite nuclear matter properties}

The nuclear matter properties of the BSkG1 and BSkG2 parameterizations are 
summarized in Table~\ref{tab:inm_prop}. We repeat that, for both models, the 
isoscalar effective mass $M_s^*/M$ and the symmetry energy $J$ were enforced 
during the parameter adjustment,
while a reasonable value for the incompressibility $K_{\nu}$ was guaranteed by 
our choice of the density-dependence parameter $\gamma = 0.3$~\cite{Chabanat97}. 
Additionally, the Fermi momentum $k_F$ was adjusted to qualitatively reproduce the global
trend of charge radii.  

The Landau parameters $G_0$ and $G'_0$ represent the effective spin-spin 
interaction between nucleons at the Fermi surface of infinite symmetric matter.
Their BSkG2 values being positive is a consequence of our finding that
the spin terms are repulsive at all densities encountered in nuclei. 
Making the same assumptions as outlined in Sec.~\ref{sec:SkyrmeEnergy} for the 
time-odd terms, we obtain very similar values of the Landau parameters $G_0$ 
and $G_0'$ for BSkG1. Including the time-odd sector in the parameter adjustment
thus did not significantly impact the Landau parameters $G_0$ and $G'_0$; 
both BSkG1 and BSkG2 values are compatible with the recommended values 
($G_0 = 0.4, G'_0 = 1.2$) 
of Ref.~\cite{Bender02}\footnote{We remind the reader that the EDF does not
include terms of the form $J^{2}_{t, \mu \nu}(\bold{r}) - \bold{s}_t(\bold{r}) \cdot \bold{T}_t(\bold{r})$, 
such that the 
Landau parameters $G_1$ and $G_1'$ vanish by construction.}. 

\begin{figure}[t!]
\includegraphics[width=.5\textwidth]{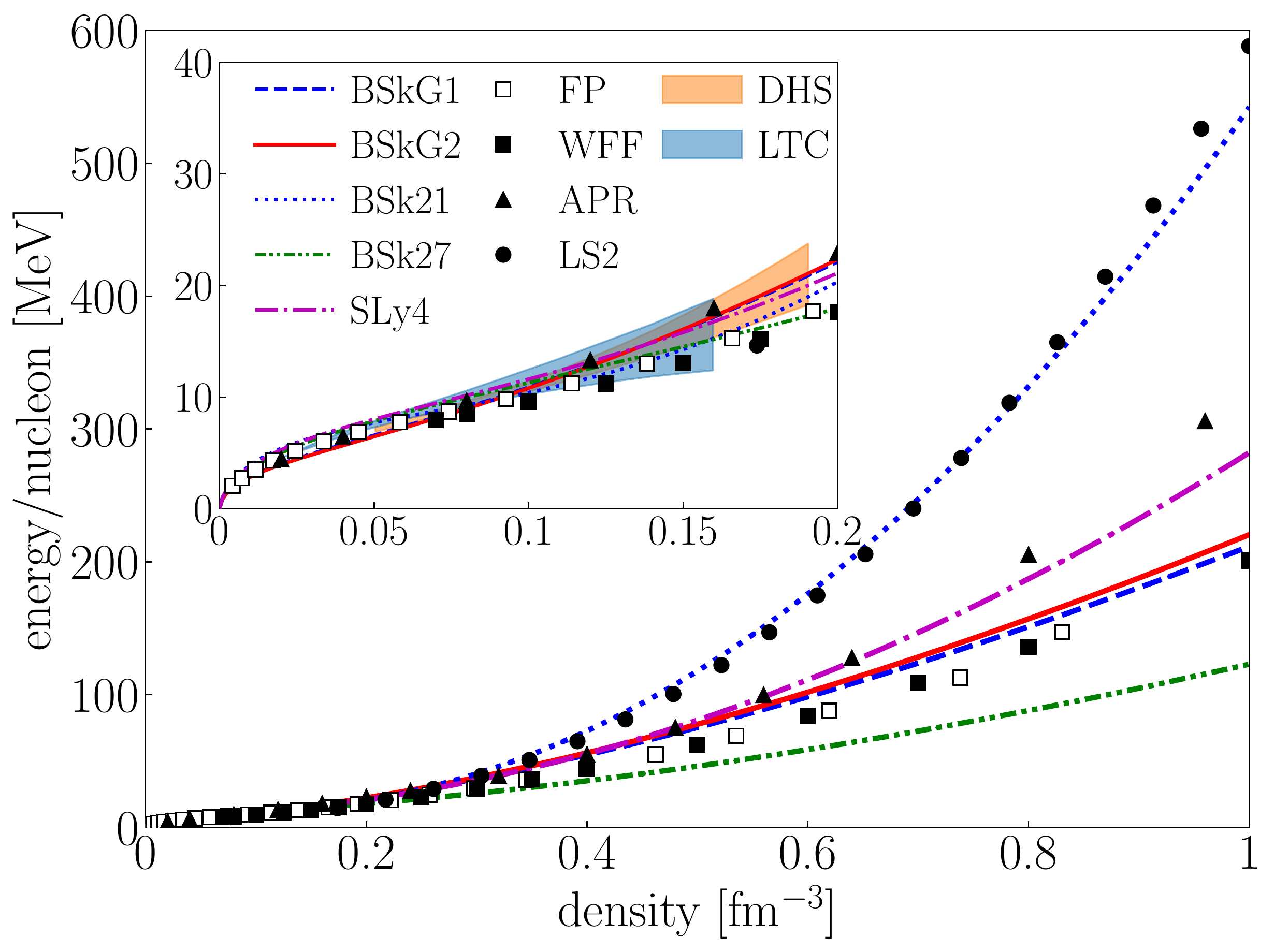}
\caption{(Color online) Zero-temperature EOS for pure neutron 
matter with BSkG2, compared to ab-initio
calculations referred to as FP, WFF, APR, LS2, DHS and LTC,  as well as the 
BSkG1~\cite{Scamps21}, BSk21~\cite{Goriely10}, BSk27~\cite{Goriely13a} and 
SLy4~\cite{Chabanat98} Skyrme parameterizations. FP corresponds to the 
calculation of Friedman \& Pandharipande \cite{Friedman81}, 
WFF to ``UV14 plus TNI'' of Ref.~\cite{Wiringa88}, 
APR of ``A18 + $\delta\,v$ + UIX$^*$" in Ref.~\cite{Akmal98} and
LS2 to V18 in Ref.~\cite{Li08}.  
The bands labelled LTC and DHS refer to predictions based on chiral 
interactions from Refs.~\cite{Lynn16} and \cite{Drischler19} respectively. 
The upper left insert is a zoom in view of densities below 0.2~fm$^{-3}$.} 
\label{fig:neutmat}
\end{figure}

In Fig.~\ref{fig:neutmat}, we compare the neutron matter equation of state (EOS)
of BSkG2 with that predicted by other EDF-based models (BSkG1, BSk27, BSk21 and SLy4)
and different types of ab-initio calculations. The BSkG2 EOS is virtually
identical to the EOS predicted by BSkG1, and only at extreme densities a
small difference is visible. This means that BSkG2 shares the qualities and flaws
of BSkG1: at low densities the models fall just below the 
uncertainty bands of the recent chiral EFT-based predictions of Ref.~\cite{Lynn16}
and Ref.~\cite{Drischler19}, see the inset of Fig.~\ref{fig:neutmat}.  
At high densities the BSkG2 EOS is moderately stiff
due to the enforced value of the symmetry energy $J=32$~MeV and qualitatively agrees with 
the WFF~\cite{Wiringa88} and FP~\cite{Friedman81} calculations, but is 
much less stiff than the LS2 prediction~\cite{Li08}. Both the BSkG1 and BSkG2
models are intermediate in stiffness between BSk27 and BSk21 and close to the 
EOS obtained with SLy4 that itself has been adjusted to reproduce the 
``UV14+UVII'' results of Ref.~\cite{Wiringa88}. We note in particular that an 
EOS of moderate stiffness is at odds with the observation of massive pulsars 
like PSR J0740+66220~\cite{Fonseca21}.

We do not discuss explicitly other properties of unpolarised infinite nuclear
matter, as BSkG1 and BSkG2 are nearly identical in these respects. In particular, 
we do not show the decomposition of the potential
energy per nucleon among the four two-body spin-isospin $(S,T)$ channels, nor the neutron
and proton effective masses in symmetric nuclear matter, as the resulting figures are 
virtually identical to Figs. 13 and 14 of Ref.~\cite{Scamps21}. 
Calculations of polarised infinite neutron matter are discussed in Sec.~\ref{sec:pnm}.


\subsection{Testing the time-odd terms}
\label{sec:testing}

Time-odd terms also impact other quantities besides masses and the 
nuclear shape. We report here on some further benchmarks of the model for 
a limited amount of observables and nuclei. Aside from calculations with 
the BSkG2 parameterization as described above, we also employ three different 
variations of BSkG2 that have identical time-even parts but
 treat the time-odd channel differently:
\begin{itemize}
\item BSkG2$_{\rm even}$: the BSkG2 parameterization with \emph{all} time-odd coupling constants set to zero.
\item BSkG2$_{\rm GI}$: the BSkG2 parameterization with all coupling constants of the spin terms set to zero, but with current and time-odd spin-orbit terms unchanged.
\item BSkG2$_{\rm double}$: the BSkG2 parameterization with all coupling constants of the spin terms doubled, but with current and time-odd spin-orbit terms unchanged.
\end{itemize}
We remark that setting all time-odd coupling constants to zero implies that
calculations with BSkG2$_{\rm even}$ are not Galilean invariant, and 
should hence be taken as purely illustrative. The two other parameterizations 
are manifestly Galilean invariant, as is BSkG2 itself.

\begin{figure}[t!]
\includegraphics[width=.48\textwidth]{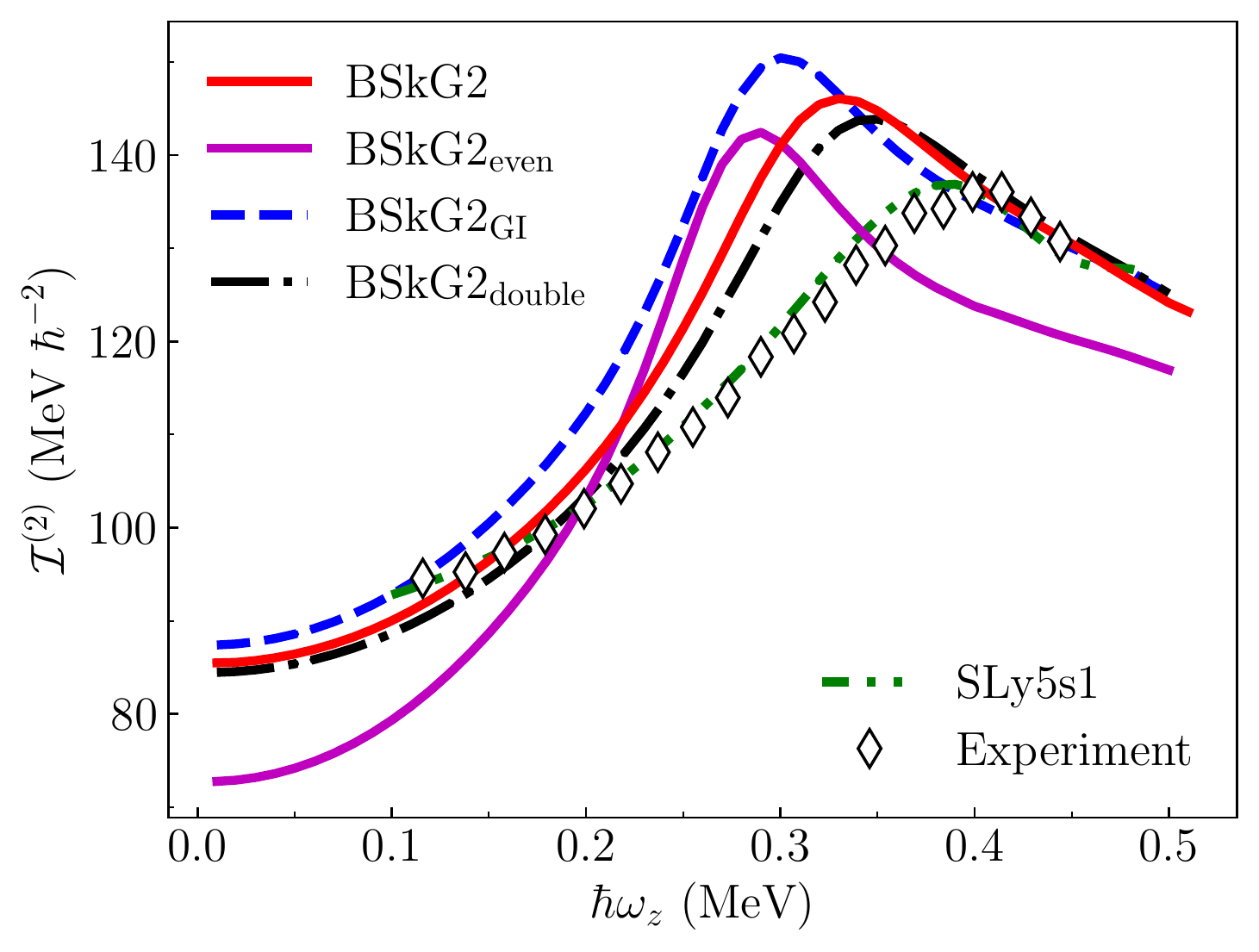}
\caption{(Color online) Dynamical moment of inertia $\mathcal{I}^{(2)}$ as a function of
         cranking frequency $\hbar \omega_z$ for the superdeformed band 
         of $^{194}$Hg as calculated with BSkG2 and three variations
         (see text). These are compared the experimental `SD-1' band
         of Ref.~\cite{Singh02} (open diamonds) and the SLy5s1 results of 
         Ref.~\cite{Ryssens19b} (green dash-dot-dotted line).
         }
\label{fig:SD_band}
\end{figure}

\subsubsection{Superdeformed rotational band: $^{194}$Hg}

As a first example, we performed self-consistent cranking calculations of the 
superdeformed yrast band of $^{194}$Hg (see Sec.~\ref{sec:mois}). In 
Fig.~\ref{fig:SD_band}, we compare BSkG2 results for the dynamical MOI
 $\mathcal{I}^{(2)}$ along this band to experimental information from 
 Ref.~\cite{Singh02} and the SLy5s1 results of Ref.~\cite{Ryssens19b}.
Note that (i) we did not employ the semivariational strategy for these cranking 
calculations, treating the collective correction perturbatively and (ii)
we employed the stabilisation procedure of Ref.~\cite{Erler08} to avoid 
the disappearance of pairing correlations.

The unmodified BSkG2 parameterization offers a fair description of the rotational
band, although it slightly underestimates the moment of inertia at low frequency
and the maximum of the curve does not occur at the same location. The 
agreement with experiment is strictly worse than that obtained with the SLy5s1 
parameterization, but we recall that (i) the pairing strengths used in 
Ref.~\cite{Ryssens19b} were optimized to describe superdeformed 
bands in this region~\cite{Rigollet99} and (ii) we did not employ the 
Lipkin-Nogami procedure that was used in Ref.~\cite{Ryssens19b}. In fact, a 
calculation with SLy5s1 and the pairing strengths of Ref.~\cite{Ryssens19b}, but 
replacing the Lipkin-Nogami method with the same stabilisation procedure as used here, results
in values for $\mathcal{I}^{(2)}$ (not shown) that are similar to those obtained with BSkG2.

Among all variants, only BSkG2$_{\rm even}$ produces results that are
dramatically different from those of BSkG2: at low spin this parameterization
severely underestimates the moment of inertia. Among the three Galilean-invariant
parameterizations the differences are smaller, but enlarging the spin coupling 
constants (i) reduces slightly the moment of inertia at low angular momentum 
and (ii) moves 
the maximum of the curve towards higher frequencies, i.e.\ delays the onset of 
the quasiparticle alignment in the band.  We note the value of the Belyaev 
MOI of the superdeformed configuration
is $\mathcal{I}^{\rm B} \approx 67 $ $\hbar^2$ MeV$^{-1}$. 
This value is about 25\% lower of the Thouless-Valatin MOI, $\mathcal{I}^{\rm TV} \approx 86$ 
$\hbar^2$ MeV$^{-1}$, which is consistent with what has been found for the 
MOI of normal-deformed states in Sect.~\ref{sect:pairing:properties}. 
The time-odd terms play a large role in this difference, as the Thouless-Valatin
MOI for BSkG2$_{\rm even}$ is only about $73$ $\hbar^2$ MeV$^{-1}$.

Finally, these calculations serve the secondary purpose
of testing the stability of the BSkG2 parameterization with respect
to spurious finite-size instabilities~\cite{Hellemans13}. The highest rotational
frequencies shown in Fig.~\ref{fig:SD_band} correspond to large angular 
momenta of about $50 \hbar$, hence to large spin and current densities. Even in
such conditions, we have seen no sign of spurious finite-size instabilities 
in the spin channels.

\begin{figure}[t!]
\centering
\includegraphics[width=.48\textwidth]{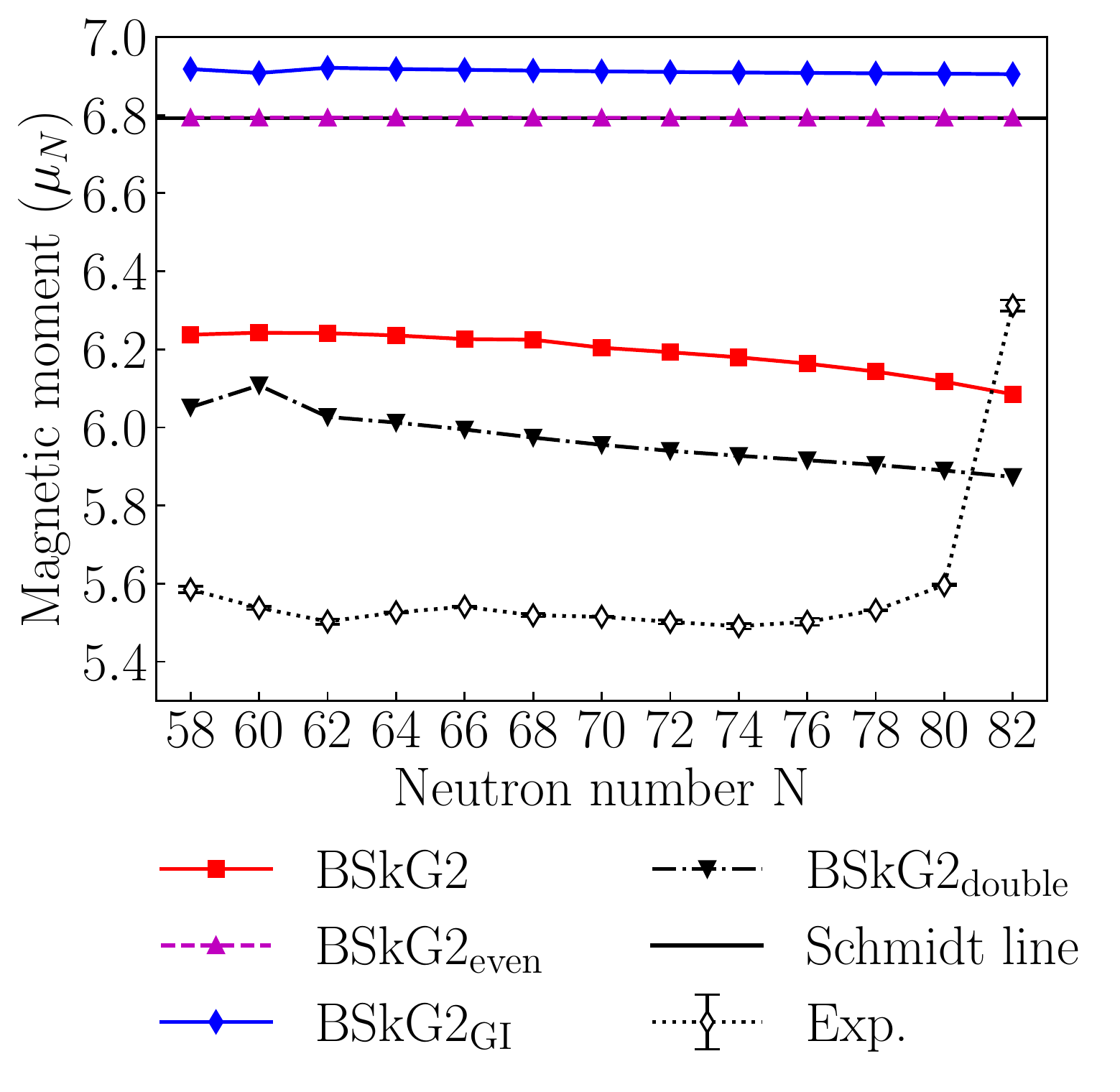}
\caption{Magnetic moments of $\tfrac{9}{2}^+$ states in odd-mass In isotopes, 
         calculated with BSkG2 and three variations (see text), as well as 
         experimental data from Refs.~\cite{Eberz87} and 
         \cite{Vernon22}. 
         \label{fig:magneticmoment}
         }
\end{figure}

\subsubsection{Magnetic moments: In isotopes}

As second example of observables impacted by the time-odd terms, we study 
the magnetic moments of a limited set of nuclei. Following Bohr and 
Mottelson~\cite{Bohr69}, the magnetic moment $\mu$ of a nucleus in a state 
with spin $J$ is defined as
\begin{align}
\mu = \langle J M=J | \hat{M}_{1z} | J M=J \rangle \, ,
\label{eq:magnetic_def}
\end{align}
where the expectation value is taken with respect to the substate with largest
possible projection of the angular momentum $M=J$ on the $z-$axis in the laboratory frame. 
$\hat{M}_{1z}$ is the $z$-component of the $M1$ operator, that is given by
\begin{align}
\hat{\boldsymbol{M}}_1 =
\sum_{i} \big( g_{\ell,i} \, \hat{\boldsymbol{\ell}}_i + g_{s,i} \, \hat{\bold{s}}_i \big) 
\, ,
\label{eq:magnetic}
\end{align}
where $\hat{\bold{\ell}}_i$ and $\hat{\bold{s}}_i$ are the one-body operators for 
orbital and spin angular momentum, respectively. The $\hat{\boldsymbol{M}}_1$
operator  is evidently time-odd and, when summed over all nucleons, probes the 
current ($\boldsymbol{\ell}$) and spin ($\boldsymbol{s}$) densities in the 
nuclear ground state. We take the bare values for the orbital and spin gyromagnetic factors of protons and neutrons:
\begin{equation}
\label{eq:g:factors}
\begin{aligned}
    g_{\ell, p} &= \mu_N,         & g_{\ell,n} &= 0 \, ,\\
    g_{s,p}     &= +5.586 \mu_N,  & g_{s, n}   &= -3.826 \mu_N \, ,
\end{aligned}
\end{equation}
where $\mu_N$ is the nuclear magneton.

For quasiparticle vacua that respect axial symmetry, it is possible to connect 
expectation values of $\hat{M}_1$ calculated in the intrinsic frame to the observed
$\mu$, a quantity in the laboratory frame~\cite{Bohr75,Bonneau15,Peru21}. For more
general symmetry-broken states, there exists to the best of our knowledge 
no simple recipe to make the connection between both frames without resorting 
to full-fledged symmetry restoration approaches~\cite{Sassarini21,Vernon22,Bally22}. 
Since our aim is an initial study of the effect of time-odd terms, we limited ourselves to one isotopic
chain of (near-)spherical isotopes. In that simple case, the magnetic moment of 
the state with spin $J$ can be calculated directly from the quasiparticle vacuum
$|\Phi\rangle$ as $\mu \approx \langle \Phi |\hat{M}_{1z} | \Phi \rangle$, provided 
this auxiliary state is constructed by blocking a quasiparticle with maximal 
alignment along the $z$-axis, i.e.\ whose angular momentum projection
on the $z$-axis is close to $J$.

Motivated by recent experimental results~\cite{Vernon22}, we selected the $J=\tfrac{9}{2}^+$ 
ground state of odd In ($Z=49$) isotopes as a testing ground. To construct near-spherical 
quasiparticle vacua, we (i) started the calculations from a carefully prepared spherically
symmetric initial guess, (ii) used the direct diagonalization approach to the HFB problem to select
a positive parity quasiparticle of the appropriate angular momentum near the 
Fermi energy at every iteration and (iii) used a cranking constraint, 
Eq.~\eqref{eq:routhian}, to keep the total angular momentum along the $z$-axis 
equal to $\frac{9}{2}\hbar$.
Finally, we did not execute a semivariational search for the minimum of $E_{\rm tot.}$,
i.e.\ we took the correction energy $E_{\rm corr}$ as a perturbation.

In Fig.~\ref{fig:magneticmoment}, we compare the magnetic moments obtained in this way
with BSkG2 and its variants to experimental data from Refs.~\cite{Vernon22,Eberz87}. 
From $N=58$ to $N=80$, the measured values show a quite strong and systematic deviation from the 
Schmidt line, which represents the magnetic moment of a non-interacting nucleon 
orbiting around a perfectly spherical core~\cite{Schmidt34}. This reduction of 
the magnetic moment is generally taken to be a sign of current and spin
distributions induced in the $Z=50$ core by the odd proton hole. The dramatic 
change of the magnetic moment for the heaviest isotope indicates that 
such polarisation is significantly less strong for a core with 82 neutrons. 

A calculation with the original BSkG2 model leads to significant core polarisation, 
although not enough to match the experimental results. This polarisation is 
completely absent in a calculation with BSkG2$_{\rm even}$, i.e.\ without time-odd terms, 
which matches exactly the Schmidt line\footnote{We repeat that BSkG2$_{\rm even}$
is not Galilean invariant, such that its properties should only
be taken as illustrations.}. Considering only the coupling constants 
of time-odd terms dictated by Galilean invariance, we see that a 
small amount of polarization appears for BSkG2$_{\rm GI}$ though with the wrong 
sign, indicating that the spin terms induce most of the core polarisation
found with the original BSkG2. Doubling the coupling constants of the spin
terms enhances the polarisation further beyond the BSkG2 results.

While not being entirely satisfying, the overall level of agreement with 
data found for BSkG2 is similar, if not better than what is typically found 
for other (unaltered) parameterisations of the nuclear 
EDF~\cite{Sassarini21,Bonneau15,Peru21,Vernon22}. To improve on the situation, it
has recently been suggested to adjust the time-odd terms of Skyrme EDFs to 
experimental data on magnetic moments~\cite{Sassarini21,Vernon22}. 
While our results in Fig.~\ref{fig:magneticmoment} confirm that varying the 
time-odd terms can resolve at least partially the tension with experiment, several 
questions remain to be solved before magnetic moments can be incorporated into
large-scale models likes ours. First: what part of the 
observed polarisation should be reproduced by a symmetry-broken mean-field 
calculation? Given the constraints of a large-scale parameter adjustment, the 
symmetry restoration techniques used in Refs.~\cite{Sassarini21,Vernon22} are 
likely out of reach for the immediate future. This question is particularly 
relevant in view of the dramatic difference between the magnetic moment 
of $^{131}$In and all other isotopes: BSkG2 and its variants produce smooth 
curves while symmetry-restored calculations with the UNEDF1 Skyrme 
parameterization are able to reproduce the discontinuity~\cite{Vernon22}. 

While probably a necessary ingredient of calculations of magnetic moments
of the majority of deformed nuclei, from the few published studies it remains difficult to 
quantify the role and impact of symmetry-restoration for the description
of the magnetic moments of spherical nuclei as the ones discussed here, 
and more analyses of the interplay of all modelling ingredients are clearly 
needed. For example, the calculations of Ref.~\cite{Sassarini21,Vernon22} 
did not include pairing correlations; their absence can be expected to enhance
core polarisation effects.

Finally, we mention that it is not entirely clear if Eq.~\eqref{eq:magnetic_def} 
represents the entire $M1$ operator, as on very general grounds
one can expect additional contributions from two-body currents \cite{Ericson88a} 
and possibly other corrections from the consistent mapping of the full-many body problem
onto an EDF approach, see the discussion in Ref.~\cite{Bally22} and
references therein. This issue is different from any empirical modification 
of the $g$-factors $g_{\ell/s}$ that could be justified by limitations of 
the model space employed. Since we treat all nucleons equally, instead of 
differentiating between a core and a valence space, we use the bare $g$-factors
in Eq.~\eqref{eq:g:factors}.

\begin{figure}[t!]
\includegraphics[width=.5\textwidth]{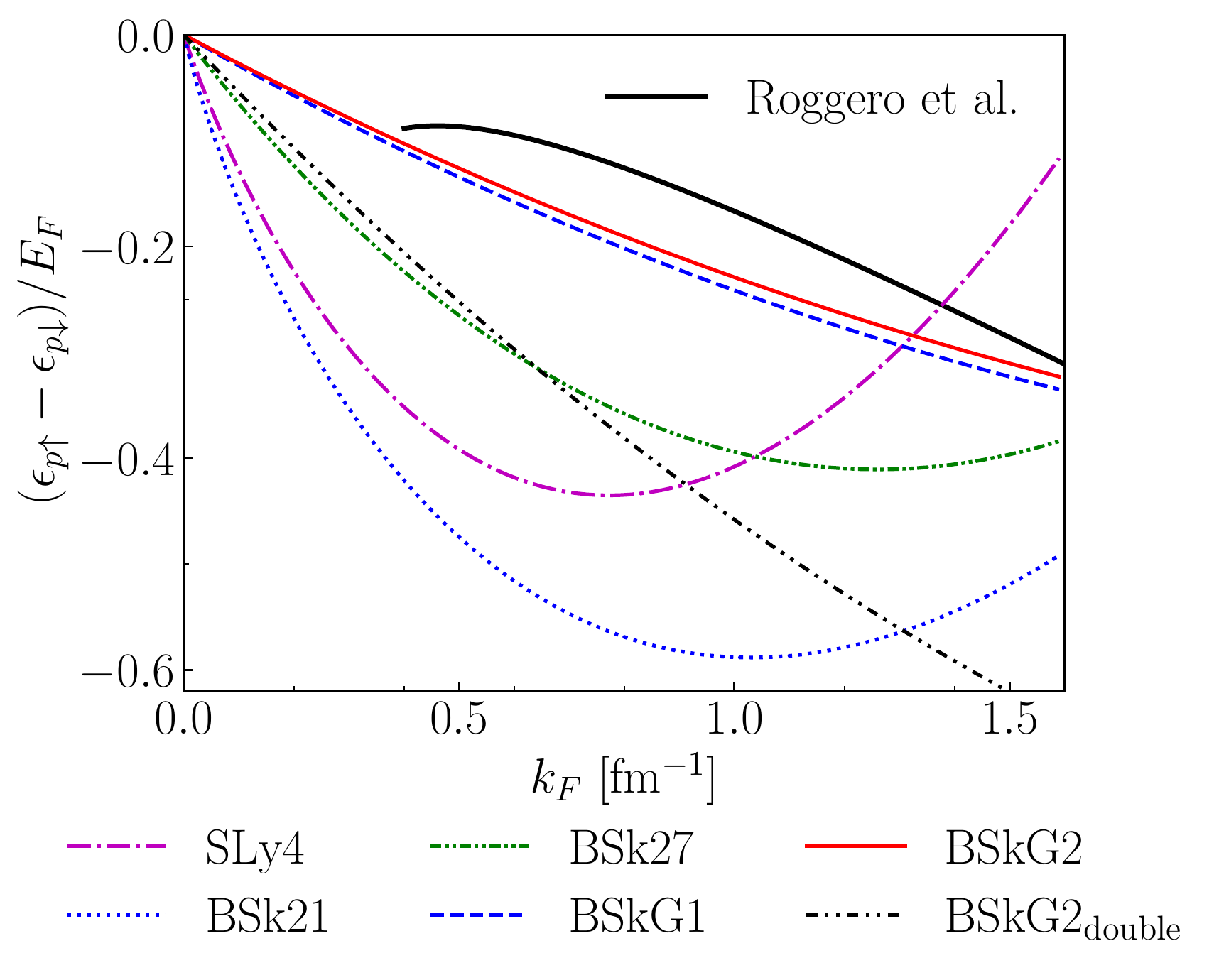}
\caption{(Color online)
Energy difference of between a spin-up and spin-down proton impurity in 
fully polarised neutron matter (see text) in units of the Fermi energy of the 
spin-up neutrons, and as a function of the Fermi momentum $k_F$. We show the predictions 
by the BSkG2 and BSkG2$_{\rm double}$ parameterization and by ab-initio calculations (full black line) 
of Ref.~\cite{Roggero15}.
For comparison, we also plot 
the predictions of the BSkG1~\cite{Scamps21}, BSk21~\cite{Goriely10}, 
BSk27~\cite{Goriely13a} and SLy4~\cite{Chabanat98} Skyrme parameterizations.
}
\label{fig:polmatter} 
\end{figure}

\subsubsection{Polarised neutron matter}
\label{sec:pnm}

All infinite matter constraints that impact the parameter adjustment 
concern \emph{unpolarized} matter, whose properties only directly impact
the time-even part of the functional. Recently, it has been suggested that 
ab-initio calculations of \emph{polarised} neutron matter can be used to 
constrain the time-odd parts of the functional as well~\cite{Roggero15}. 
A particular quantity that is affected only by the coupling constants of the 
time-odd terms in the EDF is the difference in the energy of a single
spin-up and a single spin-down proton in a sea of of spin-up neutrons. 
We compare the predictions of several Skyrme parameterizations to the 
results of the ab-initio Monte Carlo calculations in Fig.~\ref{fig:polmatter}.
Although we did not constrain specifically the time-odd part of the EDF to
reproduce this quantity, the BSkG1 and BSkG2 models are generally the closest to 
the ab-initio results. The BSkG2$_{\rm double}$ variant is wildly off and the
BSkG2$_{\rm even}$ and BSkG2$_{\rm GI}$ variants result in zero 
energy difference between spin-up and spin-down impurities (curves not shown). 
As concluded in Ref.~\cite{Roggero15}, it is clear that none of the 
parameterizations, including BSkG2, can reproduce even qualitatively the shape 
of the black full line, indicating the need to extend (the time-odd part of) the 
Skyrme EDF beyond the form we employ here. 

\section{Conclusion and outlook}
\label{sec:conclusions}

\subsection{Conclusion}

We have presented a new entry in the series of Brussels-Skyrme-on-a-Grid (BSkG)
models, which improves on the BSkG1 model of Ref.~\cite{Scamps21} in two main 
ways. First: we removed the assumption of time-reversal symmetry and 
consistently employed the entire Skyrme EDF, including the time-odd terms. 
Second, we incorporated information on the fission properties of twelve actinide
nuclei in the fitting protocol. The resulting model achieves an excellent 
description of ground-state properties 
with rms deviations of 0.668 MeV and 0.0274 fm on known masses and charge radii. 
Compared to the earlier BSkG1, the description of all observables entering the 
objective function slightly improves, such as the rms deviation of the 
AME2020 masses that is reduced by 63 keV.
The infinite nuclear matter properties of the model remain close to those of 
BSkG1. BSkG2 reproduces reference values for primary and secondary fission barriers
of 45 actinide nuclei with an rms deviation of 0.44 MeV and 0.47 MeV respectively, 
significantly improving on the BSkG1 deviations (0.88 MeV and 0.91 MeV, respectively).
To the best of our knowledge, this is the best agreement with the reference
data of RIPL-3 that is available in the literature; we will present our fission
calculations in detail in a forthcoming paper~\cite{Ryssens22}.

The parameter adjustment itself constitutes a major technical achievement: during the 
parameter adjustment we performed an enormous number of constrained Skyrme-HFB 
calculations for thousands of odd-mass and odd-odd nuclei with full inclusion
of time-reversal breaking in 3D coordinate space. We managed to eliminate a
large part of the convergence problems that plague this kind of 
calculations by using the gradient method to iterate the auxiliary states.

The changes brought by the time-odd terms to observables at fixed coupling constants
have to be distinguished from the changes brought by the presence of the time-odd terms to 
the same observables after the readjustment of the parameters. For the majority of nuclei, 
switching the time-odd terms on for a given set of coupling constants changes the masses 
of odd- and odd-nuclei by a larger amount than the actual size of the time-odd terms, 
indicating sizeable polarization effects. In the parameter adjustment the change of binding
energy brought by the presence of time-odd terms is then, however, counteracted for by 
changes of the other terms in the EDF.

For BSkG2, the total contribution of the time-odd terms to the masses of odd-mass and 
odd-odd nuclei is small (typically below 0.2 MeV in absolute value), although 
outliers for light nuclei can reach 1 MeV. These values are compatible with what has 
been observed in earlier studies, with the coupling constants of the time-odd terms 
also being fairly consistent with what is expected from phenomenology.
Including or not the time-odd terms then typically alters separation energies by 
less than 200 keV, which for weakly-bound nuclei can, however, present a large relative change.
When calculating five-point pairing gaps $\Delta^{(5)}$, including or not the time-odd terms,
results in changes by $5$ to $10\%$.
Nuclear deformation and charge radii are in general not strongly affected, although 
in exceptional situations the presence of the the time-odd terms can change the relative 
balance of coexisting minima. We have verified these 
conclusions across the entire nuclear chart, i.e.\ further than any previous
study, and have established that these effects are roughly constant up to 
the neutron drip line. All of these results also present a large-scale
benchmark of the EFA. 

Since the overall effect of time-reversal symmetry breaking remains small,
especially when compared to the typical deviation between calculations and 
experiment, the full inclusion of time-odd terms into the fitting procedure only
leads to a small improvement of masses and mass differences compared to the use of 
the EFA as in BSkG1.

The effect of time-odd terms is not necessarily small for other quantities that are 
typically not included in large-scale parameter adjustments. We have illustrated
the effect of time-odd terms on magnetic moments in In isotopes and on the 
superdeformed rotational band of $^{194}$Hg. We have also compared our results 
to {\it ab initio} results for polarised infinite nuclear matter and shown 
the difference between Belyaev and Thouless-Valatin moments of inertia.

It should be kept in mind that our study chose one specific treatment for 
the time-odd terms that does not introduce additional free coupling 
constants. Other choices are possible, although the requirement of 
Galilean invariance will always link at least some of the time-odd
terms to the time-even ones. 
Furthermore, we did not explore all possible orientations of the blocked
quasiparticles. It seems unlikely however that modifying our treatment of 
time-odd terms or exploring more general nuclear configurations would alter
any of our conclusions significantly.

\subsection{Outlook}

BSkG2 shares some of the weaknesses of BSkG1 already pointed out in Ref.~\cite{Scamps21}. 
Both models predict only a relatively soft EOS for neutron matter, which
is not compatible with the observation of pulsars of up to two solar masses. One
possibility to remedy this issue without deteriorating the description of the 
properties of finite nuclei is to adopt the extended Skyrme form of Ref.~\cite{Chamel09}. 

Secondly, our treatment of correlation energy from rotational and
vibrational motion remains schematic. Ideally, we should move beyond our 
symmetry-broken mean-field treatment and incorporate systematically symmetry
restoration and configuration mixing techniques. Doing so requires
both technical and formal developments. Technical work is necessary to 
make these techniques computationally feasible at the scale of the nuclear chart
and in particular for systems with both even and odd numbers of nucleons. 
The formal challenge consists of finding a form of the Skyrme EDF based on 
multi-nucleon interactions which can rival the performance of the standard form 
using density dependent coupling constants~\cite{Duguet09a}.  
Simpler approaches can be tried first, to be used as stepping stones towards that 
ultimate goal. One such direction would be the use of the Thouless-Valatin 
rotational MOI in the collective correction, instead of the perturbative Belyaev
MOI. With the current extension of our models to the time-odd sector, we can now
access this MOI by way of self-consistent cranking calculations.

An issue of the BSkG1 and BSkG2 models we discovered during this work is 
our systematic overestimation of three- and five-point mass differences in 
odd-$Z$ isotopic and odd-$N$ isotonic chains, despite the quality of our 
description of the same quantity in adjacent chains with an even 
number of protons or neutrons, repectively. 

Concerning the time-odd terms of the EDF: one could amend the fitting protocol 
with additional quantities that are sensitive to them, such as the magnetic 
moments and rotational MOI. Detailed preliminary studies are in our opinion 
indispensable, in particular to benchmark the size of beyond-mean-field effects. 
Time-odd observables are generally sensitive to the details of 
single-particle structure, meaning that it is not guaranteed that they can 
simultaneously be described in a global fashion using the standard form of the Skyrme 
EDF. In fact, that this is likely impossible can be deduced from
Refs.~\cite{Robledo14,Roggero15}: the traditional form fails to 
describe even qualitatively ab initio results for polarised neutron matter and 
often predicts relative spin orientations in ground states of odd-odd nuclei 
that are opposite to the empirical Gallagher-Mozkowski rules. Before attempting 
the construction of a new model, it would be desireable to establish a functional 
form that is better suited to satisfy these constraints.

\begin{acknowledgements}
This work was supported by the Fonds de la Recherche Scientifique (F.R.S.-FNRS) 
and the Fonds Wetenschappelijk Onderzoek-Vlaanderen (FWO) under the EOS 
Project nr O022818F. The present research benefited from computational resources 
made available on the Tier-1 supercomputer of the F\'ed\'eration 
Wallonie-Bruxelles, infrastructure funded by the Walloon Region under the grant 
agreement nr 1117545. The funding for G.S. from the US DOE, Office of Science, Grant No. DE-FG02- 97ER41014 is greatly appreciated.
S.G.\ and W.R.\ acknowledge financial support from the 
F.R.S.-FNRS (Belgium). Work by M.B.\ has been supported by the French Agence Nationale 
de la Recherche under grant No.\ 19-CE31-0015-01 (NEWFUN).

\end{acknowledgements}

\appendix

\section{Coupling constants of $E_{\rm Sk}$}
\label{app:couplingconstants}

The Skyrme energy density $\mathcal{E}$ of Eq.~\eqref{eq:Eskyrme} is determined by ten 
coupling constants, which are determined by the model parameters $t_{0-3}, x_{0-3}, W_0$ and $W_0'$.
The formulae relating these parameters to the coupling constants of the time-even 
part of the energy density were already presented in the appendices of Ref.~\cite{Scamps21}.
The coupling constants figuring in the time-odd part of the functional are
\begin{subequations}
\begin{alignat}{2}
C^{ss}_0               &=          -   \tfrac{1}{4}t_0 \left( \tfrac{1}{2} - x_0\right) \, ,  \\    
C^{ss}_1               &=          -   \tfrac{1}{8}t_0                                  \, ,   \\
C^{ss\rho^{\gamma}}_0  &=          -   \tfrac{1}{24}t_3 \left( \tfrac{1}{2} - x_3\right) \, ,  \\
C^{ss\rho^{\gamma}}_1  &=          -   \tfrac{1}{48}t_3\, ,  \\
C^{jj}_0               &=          - \tfrac{3}{16} t_1                                   
                                   - \tfrac{1}{4}  t_2 \left( \tfrac{5}{4} + x_2\right)\, ,       \\ 
C^{jj}_1               &=          + \tfrac{1}{8}  t_1\left( \tfrac{1}{2} + x_1\right)
                                   +               t_2 \left(\tfrac{1}{2} + x_2 \right)  \, ,       \\
C^{j \nabla s}_0    &= - \frac{W_0}{2} - \frac{W_0'}{4} \, , \\
C^{j \nabla s}_1    &= - \frac{W_0'}{4} \, .
\end{alignat}
\end{subequations}
To facilitate the analysis of the effective sign and size of the 
spin-spin interaction at the densities encountered in nuclei and nuclear 
matter, it is useful to combine $C^{ss}_t$ and $C^{ss\rho^{\gamma}}_t$ 
into a density-dependent effective coupling constant \cite{Hellemans12,Bender02}
\begin{equation}
C^{s s}_t[\rho_0]
= C^{ss}_t + C^{ss\rho^{\gamma}}_t \, \rho_0^\gamma \, .
\end{equation}

\section{Functional in the notation of Ref.~\cite{Ryssens21}}
\label{app:notation}
In Ref.~\cite{Ryssens21} a new notation for local mean-field densities was 
developed to facilitate current and future efforts to generalize the form of 
the Skyrme functional. For various reasons, the numerical implementation of 
mean-field densities in MOCCa follows this new convention, instead of the more
traditional notation of Sec.~\ref{sec:SkyrmeEnergy}. To make the link with 
the practical implementation, we include here the formulation of both 
parts of the Skyrme energy, Eqs.~\eqref{eq:Skyrme_te} and \eqref{eq:Skyrme_to}, 
as well as the pairing energy, Eq.~\eqref{eq:pairingenergy}, in the notation of 
Ref.~\cite{Ryssens21}.
\begin{align}
\label{eq:Skyrme_te_alt}
\mathcal{E}_{t, \rm e}(\bold{r})
& =  
              C^{\rho\rho}_t \, [D^{1,1}_t (\bold{r}) ]^2
            + C^{\rho\rho\rho^{\gamma}}_t [D^{1,1}_0 (\bold{r})]^\gamma\, [D^{1,1}_t (\bold{r})]^2 \nonumber \\
&           + C^{\rho\tau}_t \, D^{1,1}_t (\bold{r}) \, D^{(\nabla, \nabla)}_t (\bold{r})          \nonumber \\
&           + C^{\rho \Delta \rho}_t \, D^{1,1}_t (\bold{r}) \, \Delta D^{1,1}_t (\bold{r})        \nonumber \\
&           + C^{\rho \nabla \cdot J}_t  D^{1,1}_t (\bold{r}) \, \boldsymbol{\nabla} \cdot \bold{C}^{1,\nabla \times \sigma}_t (\bold{r}) 
\, , \\ 
\label{eq:Skyrme_to_alt}
\mathcal{E}_{t, \rm o}(\bold{r})
& =  
              C^{s s}_t \, \bold{D}^{1,\sigma}_t (\bold{r}) \cdot \bold{D}^{1,\sigma}_t    (\bold{r})
              \nonumber \\
&
            + C^{s s \rho^{\gamma}}_t  \,  [ D^{1,1}_0  (\bold{r}) ]^\gamma \, 
             \bold{D}^{1,\sigma}_t (\bold{r}) \cdot \bold{D}^{1,\sigma}_t  (\bold{r})
            \nonumber \\
&           + C^{j j}_t \, \bold{C}^{1,\nabla}_t (\bold{r}) \cdot \bold{C}^{1,\nabla}_t (\bold{r}) \nonumber \\
&           + C^{j \nabla s}_t \, \bold{C}^{1,\nabla}_t (\bold{r}) \cdot \nabla \times \bold{D}^{1,\sigma}_t (\bold{r}) 
\, , \\
\label{eq:pairingenergy_alt}
\mathcal{E}_{\rm pair} (\bold{r}) 
&= \sum_{q=p,n} g_q(\bold{r}) \, 
\Big[ \tilde{D}_q^{1,1}(\bold{r}) \, \tilde{D}_q^{1,1}(\bold{r}) 
            \nonumber \\
&
    \phantom{ \sum_{q=p,n} g_q(\bold{r}) \, \Big[ }
   +    \tilde{C}_q^{1,1}(\bold{r}) \, \tilde{C}_q^{1,1}(\bold{r}) 
       \Big] \, , 
\end{align}
where the functions $g_q(\bold{r})$ are defined in Eq.~\eqref{eq:pairingformfactor}.
This notation unambiguously separates the time-even contribution 
$\propto \tilde{D}_q^{1,1}(\bold{r}) \, \tilde{D}_q^{1,1}(\bold{r})$ to the pairing EDF
from the time-odd contribution 
$\propto \tilde{C}_q^{1,1}(\bold{r}) \, \tilde{C}_q^{1,1}(\bold{r})$.

\section{Explanation of the supplementary material}

We provide as supplementary material the file \newline
\textsf{Mass\_Table\_BSkG2.dat}, 
which contains the calculated ground state properties of all nuclei with 
$ 8 \leq Z \leq 110 $ lying between the proton and neutron drip lines. 
Its contents follows the same conventions as the supplementary material 
of Ref.~\cite{Scamps21}, but we repeat the contents of the columns in 
Table.~\ref{tab:suppl} for convenience. For clarity we mention that 
(i) we defined $M(N,Z)$ in Eq.~\eqref{eq:mass_def} as the atomic mass, while columns 3 and 4
of the file list mass \emph{excesses} and (ii) we report only the largest value 
among the three Belyaev MOIs $\mathcal{I}^{\rm B}_{x/y/z}$ in column 16.

\begin{table*}[]
\begin{tabular}{llll}
\hline
\hline
Column &  Quantity & Units & Explanation \\
\hline
1 & Z & $-$ & Proton number\\
2 & N & $-$ & Neutron number\\
3 & $M_{\rm exp}$ & MeV  & Experimental atomic mass excess \\
4 & $M_{\rm th}$ & MeV   & BSkG1 atomic mass excess \\
5 & $\Delta M$ & MeV      & $M_{\rm exp} - M_{\rm th}$\\
6 & $E_{\rm tot}$ & MeV   & Total energy, Eq.~\eqref{eq:Etot} \\
7 & $\beta_{20}$ & $-$    & \multirow{3}*{Quadrupole deformation} \\
8 & $\beta_{22}$ & $-$    & \\
9 & $\beta$  & $-$        & \\
10 & $E_{\rm rot}$ & MeV  & Rotational correction \\
11 & $\langle \Delta \rangle_n$ & MeV & Average neutron gap\\
12 & $\langle \Delta \rangle_p$ & MeV & Average proton gap\\
13 & $r_{\rm BSkG2}$ & fm  & Calculated rms charge radius\\
14 & $r_{\rm exp}$  & fm  & Experimental rms charge radius\\
15 &$\Delta r$ & fm   &  $r_{\rm exp}- r_{\rm BSkG2}$  \\
16 & $\mathcal{I}^B$ & $\hbar^2$ MeV$^{-1}$ & Calculated Belyaev MOI.\\
17 & par(p) & $-$ & Parity of proton qp. excitation\\
18 & par(n) & $-$ & Parity of neutron qp. excitation\\
\hline 
\hline
\end{tabular}
\caption{Contents of the \textsf{Mass\_Table\_BSkG2.dat} file.}
\label{tab:suppl}
\end{table*}


\newpage{\pagestyle{empty}\cleardoublepage}
\end{document}